\newcommand\apjcls{1}
\newcommand\aastexcls{2}
\newcommand\othercls{3}

\newcommand\papercls{\aastexcls}
\documentclass[tighten, times, twocolumn, trackchanges]{aastex62}

\if\papercls \apjcls
\usepackage{apjfonts}
\else\if\papercls \othercls
\usepackage{epsfig}
\fi\fi
\usepackage{ifthen}
\usepackage{natbib}
\usepackage{amssymb, amsmath}
\usepackage{appendix}
\usepackage{etoolbox}
\usepackage[T1]{fontenc}
\usepackage{paralist}

\usepackage{ragged2e}

\if\papercls \apjcls
\newcommand\aas{\ref@jnl{AAS Meeting Abstracts}}
\newcommand\dps{\ref@jnl{AAS/DPS Meeting Abstracts}}
\newcommand\maps{\ref@jnl{MAPS}}
\else\if\papercls \othercls
\usepackage{astjnlabbrev-jh}
\fi\fi

\if\papercls \aastexcls
\hypersetup{citecolor=blue, 
            linkcolor=blue, 
            menucolor=blue, 
            urlcolor=blue}  
\else
\usepackage[
bookmarks=true,           
bookmarksnumbered=true,   
colorlinks=true,          
citecolor=blue,           
linkcolor=blue,           
menucolor=blue,           
urlcolor=blue,            
linkbordercolor={0 0 1},  
pdfborder={0 0 1},
frenchlinks=true]{hyperref}
\fi

\if\papercls \othercls

\else

\fi

\providecommand{\adsurl}[1]{\href{#1}{ADS}}

\makeatletter
\patchcmd{\NAT@citex}
  {\@citea\NAT@hyper@{%
     \NAT@nmfmt{\NAT@nm}%
     \hyper@natlinkbreak{\NAT@aysep\NAT@spacechar}{\@citeb\@extra@b@citeb}%
     \NAT@date}}
  {\@citea\NAT@nmfmt{\NAT@nm}%
   \NAT@aysep\NAT@spacechar\NAT@hyper@{\NAT@date}}{}{}

\patchcmd{\NAT@citex}
  {\@citea\NAT@hyper@{%
     \NAT@nmfmt{\NAT@nm}%
     \hyper@natlinkbreak{\NAT@spacechar\NAT@@open\if*#1*\else#1\NAT@spacechar\fi}%
       {\@citeb\@extra@b@citeb}%
     \NAT@date}}
  {\@citea\NAT@nmfmt{\NAT@nm}%
   \NAT@spacechar\NAT@@open\if*#1*\else#1\NAT@spacechar\fi\NAT@hyper@{\NAT@date}}
  {}{}
\makeatother

\makeatletter
\DeclareRobustCommand{\lowcase}[1]{\@lowcase#1\@nil}
\def\@lowcase#1\@nil{\if\relax#1\relax\else\MakeLowercase{#1}\fi}
\makeatother

\DeclareSymbolFont{UPM}{U}{eur}{m}{n}
\DeclareMathSymbol{\umu}{0}{UPM}{"16}
\let\oldumu=\umu
\renewcommand\umu{\ifmmode\oldumu\else\math{\oldumu}\fi}
\newcommand\micro{\umu}
\if\papercls \othercls

\else

\fi

\let\oldsim=\sim
\renewcommand\sim{\ifmmode\oldsim\else\math{\oldsim}\fi}
\let\oldpm=\pm
\renewcommand\pm{\ifmmode\oldpm\else\math{\oldpm}\fi}
\newcommand\by{\ifmmode\times\else\math{\times}\fi}

\newbox{\wdbox}
\renewcommand\c{\setbox\wdbox=\hbox{,}\hspace{\wd\wdbox}}
\renewcommand\i{\setbox\wdbox=\hbox{i}\hspace{\wd\wdbox}}

\newcount\timect
\newcount\hourct
\newcount\minct
\newcommand\now{\timect=\time \divide\timect by 60
         \hourct=\timect \multiply\hourct by 60
         \minct=\time \advance\minct by -\hourct
         \number\timect:\ifnum \minct < 10 0\fi\number\minct}

\catcode`@=11

\newcommand\comment[1]{}

\newcommand\commenton{\catcode`\%=14}

\renewcommand\math[1]{$#1$}
\newcommand\mathshifton{\catcode`\$=3}

\let\atab=&
\newcommand\atabon{\catcode`\&=4}

\let\oldmsp=\sp
\let\oldmsb=\sb
\def\sp#1{\ifmmode
           \oldmsp{#1}%
         \else\strut\raise.85ex\hbox{\scriptsize #1}\fi}
\def\sb#1{\ifmmode
           \oldmsb{#1}%
         \else\strut\raise-.54ex\hbox{\scriptsize #1}\fi}
\newbox\@sp
\newbox\@sb
\def\sbp#1#2{\ifmmode%
           \oldmsb{#1}\oldmsp{#2}%
         \else
           \setbox\@sb=\hbox{\sb{#1}}%
           \setbox\@sp=\hbox{\sp{#2}}%
           \rlap{\copy\@sb}\copy\@sp
           \ifdim \wd\@sb >\wd\@sp
             \hskip -\wd\@sp \hskip \wd\@sb
           \fi
        \fi}
\def\msp#1{\ifmmode
           \oldmsp{#1}
         \else \math{\oldmsp{#1}}\fi}
\def\msb#1{\ifmmode
           \oldmsb{#1}
         \else \math{\oldmsb{#1}}\fi}

\def\supon{\catcode`\^=7}

\def\subon{\catcode`\_=8}

\def\supsubon{\supon \subon}

\newcommand\actcharon{\catcode`\~=13}

\newcommand\paramon{\catcode`\#=6}

\comment{And now to turn us totally on and off...}

\newcommand\reservedcharson{ \commenton  \mathshifton  \atabon  \supsubon 
                             \actcharon  \paramon}

\catcode`@=12
\reservedcharson

\if\papercls \apjcls

\else

\fi

\newcommand\HST{{\em HST}}

\newcommand\chisq{\ifmmode{\chi\sp{2}}\else\math{\chi\sp{2}}\fi}
\newcommand\redchisq{\ifmmode{ \chi\sp{2}\sb{\rm red}}
                    \else\math{\chi\sp{2}\sb{\rm red}}\fi}
\newcommand\Teq{\ifmmode{T\sb{\rm eq}}\else$T$\sb{eq}\fi}
\newcommand\Tb{\ifmmode{T\sb{\rm b}}\else$T$\sb{b}\fi}
\newcommand\mjup{\ifmmode{M\sb{\rm Jup}}\else$M$\sb{Jup}\fi}
\newcommand\rjup{\ifmmode{R\sb{\rm Jup}}\else$R$\sb{Jup}\fi}
\newcommand\msun{\ifmmode{M\sb{\odot}}\else$M\sb{\odot}$\fi}
\newcommand\rsun{\ifmmode{R\sb{\odot}}\else$R\sb{\odot}$\fi}
\newcommand\Rs{\ifmmode{R\sb{\rm s}}\else$R\sb{\rm s}$\fi}
\newcommand\mearth{\ifmmode{M\sb{\oplus}}\else$M\sb{\oplus}$\fi}
\newcommand\rearth{\ifmmode{R\sb{\oplus}}\else$R\sb{\oplus}$\fi}
\newcommand\Rp{\ifmmode{R\sb{\rm p}}\else$R\sb{\rm p}$\fi}

\newcommand\molhyd{H$\sb{2}$}

\newcommand\kms{km\;s$\sp{-1}$}

\newcommand\degree{\degr}

\newcommand\der{\ifmmode{\rm d}\else\math{\rm d}\fi}

\NoNewPageAfterKeywords

\newcommand\twoohnine{HD~209458}
\newcommand\twoohnineb{HD~209458\,b}

\newcommand\cloudy{\textsc{cloudy}}
\newcommand\angstrom{\ifmmode{\mbox{\normalfont\AA}}\else\AA\fi}
\newcommand\mgi{\ifmmode{{\rm Mg}}\else{Mg}\fi}
\newcommand\mgii{\ifmmode{{\rm Mg}^+}\else{Mg$^+$}\fi}
\newcommand\mgiii{\ifmmode{{\rm Mg}^{2+}}\else{Mg$^{2+}$}\fi}
\newcommand\fei{\ifmmode{{\rm Fe}}\else{Fe}\fi}
\newcommand\feii{\ifmmode{{\rm Fe}^+}\else{Fe$^+$}\fi}

\shorttitle{NUV Transmission Spectroscopy of HD~209458\,b}
\shortauthors{Cubillos {\em et al.}}

\begin{document}

\title{
Near-ultraviolet Transmission Spectroscopy of {\twoohnineb}: \\
Evidence of Ionized Iron Beyond the Planetary Roche Lobe
}

\author{Patricio~E.~Cubillos}
\affiliation{Space Research Institute, Austrian Academy of Sciences,
             Schmiedlstrasse 6, A-8042, Graz, Austria}
\author{Luca~Fossati}
\affiliation{Space Research Institute, Austrian Academy of Sciences,
             Schmiedlstrasse 6, A-8042, Graz, Austria}
\author{Tommi~Koskinen}
\affiliation{Lunar and Planetary Laboratory, University of Arizona,
             1629 E. University Blvd., Tucson, AZ 85721, USA}
\author{Mitchell~E.~Young}
\affiliation{Space Research Institute, Austrian Academy of Sciences,
             Schmiedlstrasse 6, A-8042, Graz, Austria}
\author{Michael~Salz}
\affiliation{Hamburger Sternwarte, Universit\"at Hamburg,
             Gojenbergsweg 112, 21029 Hamburg, Germany}
\author{Kevin~France}
\affiliation{Laboratory for Atmospheric and Space Physics,
             University of Colorado, 600 UCB, Boulder, CO 80309, USA}
\author{A.~G.~Sreejith}
\affiliation{Space Research Institute, Austrian Academy of Sciences,
             Schmiedlstrasse 6, A-8042, Graz, Austria}
\author{Carole~A.~Haswell}
\affiliation{School of Physical Sciences,
             The Open University, Walton Hall, Milton Keynes MK7 6AA, UK}

\email{patricio.cubillos@oeaw.ac.at}

\begin{abstract}

The inflated transiting hot Jupiter HD~209458\,b is one of the best
studied objects since the beginning of exoplanet characterization.
Transmission observations of this system between the mid infrared and
the far ultraviolet have revealed the signature of atomic, molecular,
and possibly aerosol species in the lower atmosphere of the planet, as
well as escaping hydrogen and metals in the upper atmosphere.
From a re-analysis of near-ultraviolet (NUV) transmission
observations of HD~209458\,b, we detect ionized iron ({\feii})
absorption in a 100 {\AA}-wide range around 2370 {\AA}, lying beyond
the planetary Roche lobe.  However, we do not detect absorption of
equally strong {\feii} lines expected to be around 2600
{\AA}.
Further, we find no evidence for absorption by neutral magnesium
({\mgi}), ionized magnesium ({\mgii}), nor neutral iron
({\fei}).  These results avoid the conflict with theoretical
models previously found
by \citet{VidalMadjarEtal2013aaHD209458bSTISnuv}, which
detected {\mgi} but did not detect {\mgii} from this same dataset.
Our results indicate
that hydrodynamic escape is strong enough to carry atoms as heavy as
iron beyond the planetary Roche lobe, even for planets less irradiated
than the extreme ultra-hot-Jupiters such as WASP-12\,b and
KELT-9\,b.
The detection of iron and non-detection of magnesium in the upper
atmosphere of HD~209458\,b can be explained by a model in which the
lower atmosphere forms (hence, sequesters) primarily magnesium-bearing
condensates, rather than iron condensates.  This is suggested by
current microphysical models.  The inextricable synergy between upper-
and lower-atmosphere properties highlights the value of combining
observations that probe both regions.

\end{abstract}

\keywords{Exoplanet atmospheres, Transmission spectroscopy, Radiative transfer}

\section{Introduction}
\label{introduction}

The early G-type star {\twoohnine} hosts what is considered to be the
prototypical hot Jupiter. The $\sim$0.7 Jupiter-mass planet has an
inflated atmosphere, an orbital period of about 3.5 days, and an
equilibrium temperature of about 1500\,K \citep{KnutsonEtal2007apjHD209458b}.  {\twoohnine} is among the
five brightest and nearest stars known to host a transiting hot
Jupiter \citep[see, e.g.,][]{EdwardsEtal2019ajARIELtargetList}. In addition, {\twoohnineb} was the first exoplanet detected
in transit \citep{CharbonneauEtal2000apjHD209458bTransit} and the
first for which an atmosphere was
detected \citep{CharbonneauEtal2002apjAtmosphereDetection}.
Consequently, {\twoohnineb} has been one of the most studied hot
Jupiters to date with transmission and emission observations obtained
from space and ground.  These observations led to the detection of
planetary atmospheric signatures ranging from the far-ultraviolet
(FUV) to infrared bands.

Shortly after the first detection of the atmosphere of {\twoohnineb}
at the position of the Na\,D
lines \citep{CharbonneauEtal2002apjAtmosphereDetection}, \citet{VidalMadjarEtal2003natHD209458bUpperAtmosphere}
reported the detection of an extended hydrogen envelope surrounding
the planet. This result was based on primary transit observations
obtained with the \textit{Hubble Space Telescope} (\textit{HST}) at
FUV wavelengths, which revealed a $\sim$10\% deep absorption in the
wings of the Ly$\alpha$ line \citep[the core of the Ly$\alpha$ line is
completely absorbed by the interstellar medium; see
also][]{BenJaffel2007apjlHD209458bHydrogen,
VidalMadjarEtal2008apjHD209458bEvaporation}. The strength of the
Ly$\alpha$ absorption signature indicated the presence of escaping
atomic hydrogen outside the planet's Roche lobe.  This first discovery
inspired the development of a number of computational models aiming at
describing the physical mechanisms of exoplanet atmospheric
escape \citep[e.g.,][]{
LammerEtal2003apjEUVatmosphericLoss,
Yelle2004icarGiantPlanetsAeronomy,
GarciaMunoz2007pssHD209458bAeronomy,
MurrayClayEtal2009apjAtmosphericEscape,
KoskinenEtal2013icarHD209458bMetalsEscapeI,
KoskinenEtal2013icarHD209458bMetalsEscapeII,
BourrierLecavelier2013aa3DhydroEscapeModels,
KislyakovaEtal2014sciHD209458bLyAlpha,
SalzEtal2015aaXUVMassLoss,
KhodachenkoEtal2017apjHD209458bLyAlpha,
CarrollNellenbackEtal2017mnrasHotPlanetaryWinds,
DebrechtEtal2018mnrasWASP12bCircumstellarGas,
KubyshkinaEtal2018apjOvercomeEnergyLimited} and the effort to look for
signatures of escape around other close-in planets and at different
wavelengths. In fact, further UV observations led to the detection of
escaping atmospheres around the hot Jupiters HD~189733\,b in the
FUV \citep[e.g.,][]{LecavelierEtal2012aaHD189733bEvapVariation},
WASP-12\,b in the near-ultraviolet \citep[NUV;
e.g.,][]{FossatiEtal2010apjWASP12bMetals,
HaswellEtal2012apjWASP12bNUV}, and indications for WASP-121\,b in the
NUV \citep[][]{SalzEtal2019aaWASP121bSwift}, and the warm Neptunes
GJ\,436\,b and GJ\,3470\,b in the
FUV \citep[e.g.,][]{EhrenreichEtal2015natGJ436bHydrogenEscape,
BourrierEtal2018aaGJ3470b}.

The hot Jupiter {\twoohnineb} has since been observed at both FUV and
NUV wavelengths, to further study atmospheric escape. The FUV
observations focused mostly on detecting planetary atmospheric
absorption at the position of resonance lines of abundant
elements, particularly C, O, and Si. The observations led to the
detection of O and C$^+$ escaping from the
planet \citep{VidalMadjarEtal2004apjHD209458bOyxygenCarbon,
LinskyEtal2010apjHD209458bMassLoss,
BallesterBenJaffel2015apjHD209458bHSTfuv}, while the detection of
Si$^{2+}$ is still
controversial \citep{LinskyEtal2010apjHD209458bMassLoss,
BallesterBenJaffel2015apjHD209458bHSTfuv}.

\subsection{NUV Atmospheric Characterization of {\twoohnineb}}

\citet{VidalMadjarEtal2013aaHD209458bSTISnuv} presented the
results of three {\HST} transit observations conducted at NUV
wavelengths. Observing exoplanetary atmospheric escape in the NUV has
some key advantages over the FUV \citep{HaswellEtal2012apjWASP12bNUV,
FossatiEtal2015asslExoAtmObservations,
Haswell2017HandbookOfExoplanets}: 1) for K-type stars and earlier, the
NUV stellar emission is dominated by the photospheric
continuum, which has a
significantly higher flux than that the FUV emission lines, thus
improving the signal-to-noise (S/N); 2) Because it is dominated by
photospheric emission, the background over which a transit is observed
is more homogeneous and less affected by bright/dark spots typical of
the FUV chromospheric emission \citep[see
e.g.,][]{Haswell2010bookTransitingExoplanets,
LlamaShkolnik2015apjXUVtransits, LlamaShkolnik2016apjLyAlphaTransits}.
In addition, the NUV spectral range contains a large number of resonance
lines of abundant metals (e.g., Mg, Fe, and Mn), some of which have been
detected in escaping exoplanetary
atmospheres \citep[e.g.,][]{FossatiEtal2010apjWASP12bMetals,
HaswellEtal2012apjWASP12bNUV, SingEtal2019ajWASP121bTransmissionNUV}.

{\citet{VidalMadjarEtal2013aaHD209458bSTISnuv}} reported the detection
of planetary {\mgi} absorption at $2.1\sigma$ level with Doppler
velocities ranging between $-62$ and $-19$\,{\kms} away from the
position of the {\mgi} $2853\,\angstrom$ resonance line. 
\citet{VidalMadjarEtal2013aaHD209458bSTISnuv}
and \citet{BourrierEtal2014aaHD209458bMgIIescapeModel,
BourrierEtal2015aaMgIline} interpreted this as evidence of planetary
{\mgi} atoms escaping the planet's atmosphere and moving away from its
host star, accelerated by radiation pressure.
\citet{VidalMadjarEtal2013aaHD209458bSTISnuv} also
looked for {\mgii} absorption at the position of the {\mgii}\,h\&k
resonance lines at about 2800\,\AA, but without success. In addition,
they binned the whole available spectral window to $200\,\angstrom$
bins looking for the rise of the planetary radius with decreasing
wavelength due to Rayleigh scattering, but the data quality was not
high enough to constrain the Rayleigh slope.

The low ionization potential of {\mgi} (7.65\,eV) implies that
radiation at wavelengths shorter than $\sim$1621\,{\angstrom} is
capable of ionizing the atom.  Since the photospheric continuum of the
G-type star {\twoohnine} starts rising near
1450--1500\,{\AA} \citep{VidalMadjarEtal2004apjHD209458bOyxygenCarbon,
FranceEtal2010apjHD209458bFUV}, the detection of {\mgi} and not of
{\mgii} came as a great
surprise. \citet{VidalMadjarEtal2013aaHD209458bSTISnuv} explained this
as due to electron recombination depleting {\mgii} and enhancing
{\mgi} in the atmosphere and derived a minimum electron density of
10$^{8-9}$\,cm$^{-3}$ required to achieve this.  This electron density
is about a factor of ten higher than the peak electron density
predicted by upper atmosphere models \citep[10$^{7-8}$\,cm$^{-3}$;][]{
KoskinenEtal2013icarHD209458bMetalsEscapeI,
KoskinenEtal2013icarHD209458bMetalsEscapeII,
LavvasEtal2014apjElectronDensity}.

Furthermore, {\citet{BourrierEtal2014aaHD209458bMgIIescapeModel}}
found that the best fit electron density required to match the
observations is even higher, at about 10$^{10}$\,cm$^{-3}$.  In
addition to being significantly higher than theoretical predictions,
their estimate of the {\mgii} recombination rate seems to assume that
all available electrons are used to recombine {\mgii}.  Since this is
unrealistic, we infer that for recombination to explain the detection
of {\mgi} and the non-detection of {\mgii}, the required electron
density must therefore be even larger than 10$^{10}$\,cm$^{-3}$.

Motivated by these considerations, we present a re-analysis
of the {\HST} NUV observations of {\twoohnineb}.  The primary aims of
this work are to re-investigate the detection and
non-detection of the {\mgi} and {\mgii} absorption, respectively,
search for metal absorption lines in the entire STIS NUV spectrum, and
attempt to reconcile the observations with theory.  In
Section \ref{sec:observations}, we describe the observation in more
detail.  In Section \ref{sec:analysis}, we describe our data-analysis
methodology.  In Section \ref{sec:atmosphere}, we provide a
theoretical interpretation of the observations.  In
Section \ref{sec:results} we discuss the main implications of our
analysis and compare our results to those
of \citet{VidalMadjarEtal2013aaHD209458bSTISnuv}.  Finally, we
summarize our conclusions in Section \ref{sec:conclusions}.

\section{Observations}
\label{sec:observations}

We re-analyze the three archival NUV transmission observations of the
planetary system {\twoohnine}, obtained with {\HST}, using the Space
Telescope Imaging Spectrograph \citep[STIS, program \#11576,][]{VidalMadjarEtal2013aaHD209458bSTISnuv}, the
NUV Multi-Anode Microchannel Array (NUV-MAMA) detector, and E230M
grating.  Each transit observation (one {\HST} visit) consists of
five consecutive {\HST} orbits, where each {\HST} orbit lasts for
about 90 minutes.  For each visit, the first two and last {\HST}
orbits occur out of transit, the third orbit during the transit
ingress, and the fourth orbit shortly after the mid-transit time.
Each orbit consists of 10 consecutive exposures (frames) of 200
seconds each, except for the first orbit, which consists of nine
frames to accommodate acquisition observations.  Each frame consists
of an echelle spectrum comprising 23 orders, in which each order
consists of 1024 wavelength samples, with a resolving power of $\lambda/\Delta\lambda = 30\,000$, and approximately two pixels per resolution element (where the resolution element is approximately 10 {\kms}, or equivalently, 0.09 \AA).  The entire STIS spectrum covers
the 2300--3100 {\AA} range, with some overlap between the orders.

\section{Data Analysis}
\label{sec:analysis}

We follow the standard procedure for extracting the planetary transit
signature from time-series observations.  We start the analysis from the
CALSTIS\footnote{
\href{http://www.stsci.edu/hst/stis/software/analyzing/calibration/pipe\_soft\_hist/intro.html}
{http://www.stsci.edu/hst/stis/software/analyzing/calibration/pipe\_soft\_}
\href{http://www.stsci.edu/hst/stis/software/analyzing/calibration/pipe\_soft\_hist/intro.html}
{hist/intro.html}} (version 3.4) reduced and extracted
spectra, which have been corrected for flat-field, extracted,
corrected for background emission, and wavelength and flux calibrated.
Our analysis is based on the flux calibrated spectra and their
uncertainties, which account for photon noise and uncertainties on
flat-fielding and flux calibration.  We exclude the first orbit from
each visit and the first frame from each orbit, since they show severe
systematics that we were not able to correct for.

In the following sections we describe our analysis, in which we adopt
a two-step approach.  First, we characterize instrumental systematics
by using wavelength-integrated data (white-light analysis).  Second,
we obtain wavelength-resolved transmission spectra from the data that
have been corrected for systematics (spectral analysis).

\subsection{White-light Analysis}
\label{sec:white}

In this step, we integrate the flux over each echelle spectral order
to increase the signal-to-noise ratio (S/N), and then detrend the
astrophysical signal from the instrumental-systematics.  We begin by
masking bad data points based on three criteria: first, as stated
above, we exclude the first orbit from each visit and the first frame
from each orbit; second, we exclude data points with abnormally low
uncertainties, i.e., an order of magnitude lower than the median
uncertainties (typically, a handful of data points at the edges of the
echelle order for each frame); and third, we use the overlap between
the echelle orders to exclude data points where the fluxes differ from
each other by more than $5\sigma$.  Overall, this bad-pixel-masking
procedure removes typically five pixels at each end of the echelle
order, for each frame.  We then produce the raw white light curves by
summing the flux over each spectral order and propagating the
uncertainties accordingly.

We fit the raw light curves with parametric transit and systematics
models as a function of time ($t$) and wavelength ($\lambda$):
\begin{equation}
F(t, \lambda) = F\sb{\rm s} T(t, \lambda) S(t, \lambda),
\end{equation}
where $F\sb{\rm s}$ is the out of transit flux, $T(t, \lambda)$ is
a \citet{MandelAgol2002apjLightcurves} transit model, and
$S(t, \lambda)$ is a model of the instrumental systematics.
To obtain statistically robust parameter estimations, we apply a
Levenberg-Marquardt optimization to compute the best fit parameters,
and a Markov-chain Monte Carlo algorithm to compute the parameter
credible intervals.  To this end, we use the open-source MC3
Python package\footnote{\href{https://mc3.readthedocs.io/}
{https://mc3.readthedocs.io/}} \citep{CubillosEtal2017apjRednoise}.
For the MCMC exploration, we select the Snooker Differential-evolution
MCMC sampler \citep[][]{terBraak2008SnookerDEMC}.

Except for the transit epoch, we fix the orbital parameters to the values measured at
optical wavelengths, since they are more precise than those we can
infer from the NUV data.  We assume an orbital period of $P=3.5247486$ days
\citep{KnutsonEtal2007apjHD209458b}, an orbital inclination
of $i=86.66\degree$, and a semi-major axis to stellar radius ratio of
$a/R\sb{\rm s}=8.7947$ \citep{HayekEtal2012aaHD209bHD189b}.  These
values fit the NUV data well.  Given the gaps in the {\HST}
observations, adopting different orbital-parameter values from the
literature does not significantly impact our results as they impact
mostly the timing of the transit light curve rather than the relative
depths as a function of wavelength, \citep[see, e.g., furhter
discussion in][]{EvansEtal2018ajWASP121b}.

For each order, we compute the four-coefficient non-linear
limb-darkening coefficients \citep[as defined
in][]{Claret2000aaLimbDarkening} using the open-source routines
of \citet{EspinozaJordan2015mnrasLimbDarkeningI}, which we keep fixed
during the fit.  For this calculation, we consider the PHOENIX stellar
model that most closely matches the physical properties of
{\twoohnine} (effective temperature $T\sb{\rm eff}=6000$~K, surface
gravity $\log g=4.5$, and solar elemental metallicity).  Although
fixing the limb-darkening coefficients (instead of fitting) is not
optimal, the low S/N and the sparse coverage of the NUV transit light
curve does not allow us to constrain the coefficients sufficiently
well.  Therefore, the transit depth and out-of-transit flux remain the
only astrophysical fitting parameters.

In addition to the astrophysical signal, we simultaneously fit the
{\HST} instrumental systematics.  {\HST} time-series observations show
well known systematics that affect the different instruments and
observing modes on board
\citep[e.g.,][]{WakefordEtal2016apjHSTsystematics,
AlamEtal2018ajWASP52bTransmission,
SingEtal2019ajWASP121bTransmissionNUV}: the ``breathing'' effect,
which varies with the {\HST} orbital period, and visit-long
trends.  We model these systematics
considering a family of polynomial curves as
in \citet{VidalMadjarEtal2013aaHD209458bSTISnuv}:
\begin{eqnarray}
\nonumber
S(t, \phi) &= 1& +\ a\sb{1}(t-t\sb{0}) + a\sb{2}(t-t\sb{0})\sp{2} \\
\nonumber
  &&  +\ b\sb{1}(\phi-\phi\sb{0})       + b\sb{2}(\phi-\phi\sb{0})\sp{2} \\
  &&  +\ b\sb{3}(\phi-\phi\sb{0})\sp{3} + b\sb{4}(\phi-\phi\sb{0})\sp{4},
\label{eq:ramp}
\end{eqnarray}
where $\phi$ are the {\HST} orbital phases; $a\sb{i}$ and $b\sb{i}$
are fitting parameters for the visit-long and breathing systematics,
respectively; and $t\sb{0}$ and $\phi\sb{0}$ are a fixed reference
time and phase.  We set $t\sb{0}=T\sb{0}(t)$ and $\phi\sb{0}=0.2$, the
transit mid-time and {\HST} mid-phase of the observations, respectively, to minimize
correlations between the polynomial coefficients. The expression shown
in Eq. (\ref{eq:ramp}) is the most complex form of $R(t, \phi)$
we considered, as we tested all possible lower-order polynomial
expressions in $t$ and $\phi$.

We fit the light curves testing each combination of polynomial degree in
$t$ and $\phi$, selecting the preferred model by minimizing the
Bayesian Information Criterion: ${\rm BIC} = \chi\sp{2} + k\log N$,
where $k$ is the number of free parameters and $N$ is the number of
data points.  Table \ref{table:bic} shows the results of the Bayesian
model selection.  For all visits, the BIC prefers linear
polynomials in $t$ and $\phi$.

{\renewcommand{\arraystretch}{1.0}
\begin{table}[ht]
\centering
\caption{BIC Model Comparison}
\label{table:bic}
\begin{tabular*}{\linewidth} {@{\extracolsep{\fill}} cccc}
\hline
\hline
Polynomial degree &              & BIC          &         \\
($t$, $\phi$)     & Visit 1      & Visit 2      & Visit 3 \\
\hline
1, 1              & {\bf 1495.2} & {\bf 1952.0} & {\bf 2164.4}  \\
1, 2              & 1546.0       & 2042.9       & 2301.0  \\
1, 3              & 1656.4       & 2142.5       & 2276.7  \\
1, 4              & 1783.9       & 2282.5       & 2377.6  \\
2, 1              & 1623.6       & 1974.4       & 2196.6  \\
2, 2              & 1674.1       & 2053.5       & 2334.0  \\
2, 3              & 1784.6       & 2147.8       & 2308.9  \\
2, 4              & 1912.0       & 2287.7       & 2408.5  \\
\hline
\end{tabular*}
\end{table}
}

\begin{figure}[tb]
\centering
\includegraphics[width=\linewidth, clip, trim=10 0 0 0]{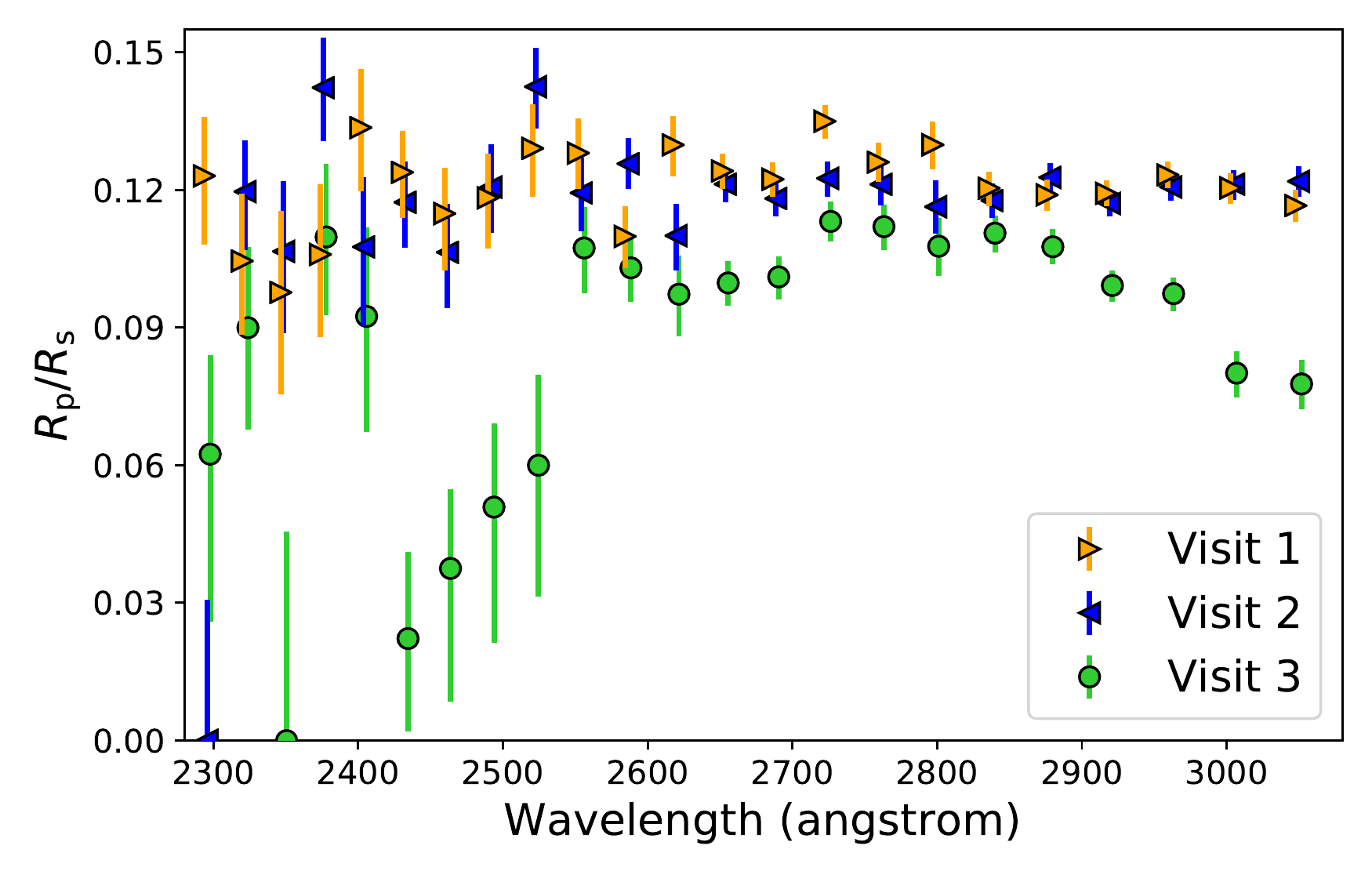}
\caption{
{\twoohnineb} white-analysis transmission spectra.  Each data point
shows the best-fitting systematics-corrected radius ratio
for the echelle orders of each visit (see legend).  The error
bars denote the 68\% highest-posterior-density credible region
(1$\sigma$ uncertainty).}
\label{fig:white}
\end{figure}

Figure \ref{fig:white} shows the best-fitting transit depth for each
visit and spectral order from the white-light analysis, for the
preferred systematics model in Table \ref{table:bic}.
Figure \ref{fig:appendix_lc} shows the order-by-order white light
curves and their best fitting transit
model.  Figure \ref{fig:appendix_systmatics} compares sample
order-integrated light curves with and without the ramp systematic
correction.  As in \citet{VidalMadjarEtal2013aaHD209458bSTISnuv}, the
transit depths of the first two visits agree well with each other and
show a nearly constant transit depth with wavelength.  In contrast,
the third visit shows strong anomalous systematics of unknown origin
that corrupt the transit fit, producing wide variations in transit
depth across the spectrum.  Thus, as done
by \citet{VidalMadjarEtal2013aaHD209458bSTISnuv}, we discard the
results from the third visit from further analyses.

\subsection{Spectral Analysis}
\label{sec:spectral}

In this step, we divide the data into several spectral channels to
search for wavelength-dependent features in the transmission spectra.
The main complication of the analysis is choosing the appropriate
width of the wavelength channels.  On the one hand, the expected NUV
features are narrow absorption lines a few {\kms} wide, but narrow
wavelength channels do not provide sufficient S/N to distinguish such
features.  On the other hand, broader channels build up S/N, but
dilute the signal of potential spectral features.  We thus analyze the
data at four different spectral resolutions (25, 50, 100, and
1000 {\kms}) to assess this trade-off.  A posteriori, we
found that the data show isolated absorption features that get quickly
diluted as one lowers the resolution.  Therefore, our main conclusions
rely on the highest-resolution analysis at 25 {\kms}, aided by the
coarse 1000 {\kms} analysis as a continuum.

\begin{figure*}[t]
\centering
\includegraphics[width=0.95\linewidth, clip, trim=0 50 0 90]{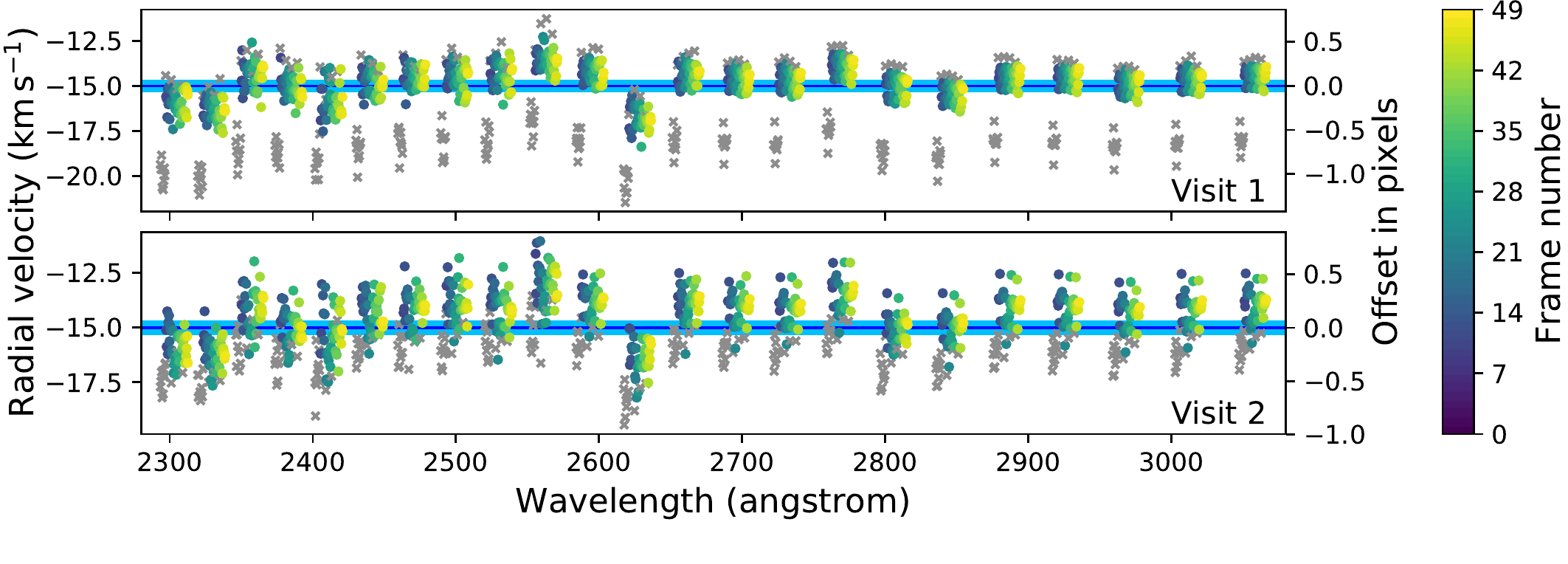}
\includegraphics[width=0.95\linewidth, clip, trim=0 50 0 90]{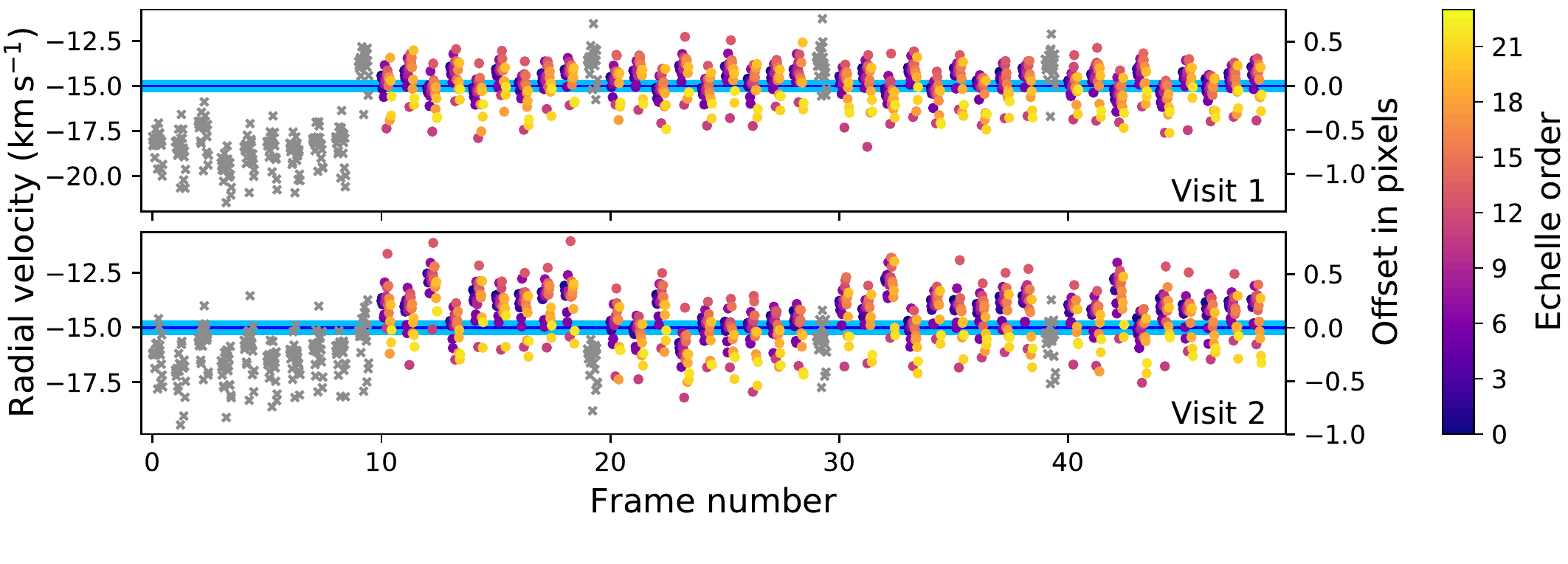}
\caption{
Wavelength calibration of each observed spectra as a function of
wavelength (top two panels) and as a function of frame number
(bottom two panels) for the first and second {\HST} visit.  The colored dots
denote the valid frames used for the fit, color coded by frame number (top two panels) and by spectral order (bottom two panels); the gray
crosses denote the discarded frames (first {\HST} orbit, and first
frame of each {\HST} orbit). The horizontal dark and light blue lines
denote the systemic radial velocity of $-15.01\pm 0.29$
{\kms} \citep{GaiaCollaboration2018catDR2}.  The markers have been
slightly shifted horizontally for better visualization.}
\label{fig:wlcal}
\end{figure*}

\subsubsection{Wavelength Correction}

We begin the spectral analysis by reading and masking the raw light
curves as in the white-light analysis.  To account for possible
wavelength distortions of the spectra during a visit, we calibrate and
correct the wavelength solution of each echelle order in each frame,
with respect to the stellar rest frame.  As a reference spectrum, we
use a high-resolution line-by-line stellar photospheric emission model
of {\twoohnine}, computed with the 1D plane-parallel stellar
atmosphere code of \citet{ShulyakEtal2004aaLBLstellarModels}.  We find
the wavelength correction for each order from the peak of the
cross-correlation function between the Doppler-shifted data and the
reference stellar spectrum.

Once calibrated, we Doppler-shift all light-curve frames of a given
echelle order to a common wavelength solution using a cubic spline interpolation, and renormalizing the flux level to conserve the total
flux in each frame.  Since the flux uncertainties are Poison-noise dominated, we also use a cubic spline to estimate the uncertainties.  To minimize the impact of the interpolation, we select as a reference point the frame that minimized the geometric mean of the shifts (of all the frames in a given order).  As a result, 68\% of the shifts are smaller than an eighth of a pixel.  We remark that the wavelength correction is done 
order-by-order, thus it does not affect the total flux of each order, 
though the interpolation modifies the flux in each pixel, thus it 
affects the light curves of the narrow spectral channels.
Figure \ref{fig:wlcal} shows the relative Doppler shift between the
individual orders for each frame and visit.  The wavelength
calibration shows significant shifts of nearly
one pixel during a visit that varies systematically with the visit-long
frame number, {\HST}-orbit-long frame number, and echelle order.
Furthermore, each visit shows a distinct Doppler-shift pattern.

\subsubsection{Divide-white Systematics Correction}

Following \citet{KreidbergEtal2014natCloudsGJ1214b}, we adapted their
{\HST}/WFC3 ``divide-white'' spectral analysis to the {\HST}/STIS NUV
dataset.  This is a two-step approach that uses the results from the
white-light analysis to remove the instrumental systematics from the
raw spectral light curves (hence, the ``divide-white'').
Here, we construct a non-parametric systematics model by dividing the
white-light curve by its best-fit order-by-order and
visit-by-visit transit model, and normalizing the resulting quotient.
Assuming that the instrumental systematics vary weakly with
wavelength \citep[an appropriate assumption for {\HST}
instruments,][]{AlamEtal2018ajWASP52bTransmission}, we divide the
systematics model from each spectral light curve.  The resulting light
curves should then be dominated by the astrophysical transit signal.

We estimate uncertainties of the best-fit white transit model from the
standard deviation of the distribution for the white-light transit
model, generated from the white-analysis posterior distribution.  Then
we use the error-propagation formula \citep[see,
e.g.,][]{BevingtonRobinson2003DataReduction} to account for all
uncertainties throughout the steps involved to construct the
non-parametric systematics model and obtain the systematics-corrected
light curve.

Once we have obtained the wavelength- and systematics-corrected light
curves, we bin the data to a wavelength array with a constant
resolution, where for each spectral channel, we co-add all flux
contributions from the two visits and from all echelle orders (in
regions of overlapping orders), and propagate the errors accordingly.

\subsubsection{Transmission-spectrum Extraction}
\label{sec:spectralfit}

To derive robust estimates of the transmission spectrum, we apply two
independent approaches to analyze the spectral light curves.  In our
first approach, we proceed in a similar manner as for the white-light
analysis: we fit a \citet{MandelAgol2002apjLightcurves} transit model
to the systematics-corrected spectral light-curves in an MCMC run.  As
previously, we fix the transit epoch, the orbital
inclination and ratio between the orbital semi-major axis and stellar
radius ($a/R\sb{\rm s}$).  We also compute and fix the non-linear limb-darkening
coefficients for each spectral channel.  Therefore, we fit for
the transit depth and out-of-transit flux level in each spectral
channel.  Since the light curve combines data from different epochs,
we fit the out-of-transit flux level of each visit with an individual
free parameter.  This approach produces statistically robust
transit-depth estimates and maximizes the S/N; however, the results
are influenced by the assumed limb-darkening coefficients.

As a second approach, we replicate the procedure
of \citet{VidalMadjarEtal2013aaHD209458bSTISnuv}, computing the
absorption depth ($AD$) from the ratio between the total flux of the
in-transit and out-of-transit {\HST} orbits:
\begin{equation}
AD(\lambda) = 1 - \frac{F\sb{3}-F\sb{4}}{F\sb{2}-F\sb{5}},
\label{eq:ad}
\end{equation}
where $F\sb{i}$ represents systematics-corrected flux summed over the
$i$-th {\HST} orbit, from both visits, within a spectral channel at
wavelength $\lambda$.  We propagate the errors according to
Eq.~(\ref{eq:ad}) to obtain the uncertainties, as done
in \citet{VidalMadjarEtal2013aaHD209458bSTISnuv}.

The transit depths from the $AD(\lambda)$ and light-curve fit analyses
should be similar, although we expect systematic variations.  Since
some data points in $F\sb{3}$ and $F\sb{4}$ occur during transit
ingress and egress, the $AD$ depths are expected to be lower than the
light-curve fit depths.  Additionally, the $AD$ analysis does not
account for stellar limb darkening.  Nevertheless, we find a good
agreement between the transit depths obtained from the light-curve fit
and the $AD$ method (see, e.g., Figure \ref{fig:magnesium}).  Both
methods produce similar uncertainties, and their transit-depth values
agree within $1\sigma$ of each other for most spectral channels.

\section{Atmospheric Signatures}
\label{sec:atmosphere}

Both spectral analyses of the {\twoohnineb} NUV observations suggest
that the planet has a nearly flat transmission spectrum.
The high-resolution analyses show scattered absorption features across the
spectrum; however, the limited S/N of the data complicates the
detection of individual features.  Here, we describe the analysis
carried out to detect atmospheric features and discuss their physical
interpretation.

\subsection{Identification of Absorption Lines}
\label{sec:lineid}

The high-resolution transmission spectrum of {\twoohnineb} (i.e., at
25\,{\kms}) shows a large number of data points that deviate
significantly above the continuum (i.e., the low-resolution spectrum
at 1000\,{\kms}).  While from a normal distribution of 3600 data
points one expects $\sim$10 outliers deviating more than $3\sigma$
above and below the mean, we typically
find $\sim$25 data
points deviating above the continuum by more than three times
their uncertainty, suggesting that most of them are likely absorption
features rising from the planetary atmosphere.

To determine whether we can associate these outliers to a particular
atomic absorption line, we apply the probabilistic approach
of \citet{HaswellEtal2012apjWASP12bNUV}.  Using the VALD database for
the atomic line transitions \citep{PiskunovEtal1995aapsVALDdatabase}
and adopting solar elemental
abundances \citep{AsplundEtal2009araSolarComposition}, we collect all
known atomic line transitions within a given wavelength range
($\Delta\lambda$) around each candidate absorption feature (i.e.,
outlier) lying at wavelength $\lambda\sb{0}$.  To each known line
transition $i$ at wavelength $\lambda\sb{i}$ we assign a ``match''
probability based on their proximity to $\lambda\sb{0}$, elemental
abundance ($n\sb{j}$), and line strength (determined by the values of
the transition probability $g\!f\sb{i}$, by the energy of the lower
level $E\sb{\rm low}\sp{i}$ (in cm$^{-1}$) of the transition, and by
the temperature of the atmospheric layers probed by the observations
$T$) as
\begin{equation}
P'\sb{i} = \frac{|\lambda\sb{i}-\lambda\sb{0}|}{\Delta\lambda}
           \,n\sb{j}\,g\!f\sb{i}
           \,\exp(\frac{-hc E\sb{\rm low}\sp{i}}{kT}),
\end{equation}
where $h$ and $k$ are respectively Planck's and Boltzmann's constants,
and $c$ is the speed of light.  Finally, we normalize all match
probabilities to obtain a discrete probability distribution:
$P\sb{i} \equiv P'\sb{i}/\sum P'\sb{i}$.  Thus, by finding the line
transitions with a high match probability ($P\sb{i} > 68.3$\%, in this
case), this framework reveals how likely it is for a specific line
transition to be the unambiguous source of one of the $3\sigma$
outliers.  Here, we consider a window of $\Delta\lambda =
2.5\,\angstrom$, which corresponds to a velocity of about 275\,{\kms},
and adopt an atmospheric temperature of $T=10\,000$~K
\citep[e.g.,][]{KoskinenEtal2013icarHD209458bMetalsEscapeI}.

\begin{figure*}[p]
\centering
\includegraphics[width=0.92\linewidth, clip]{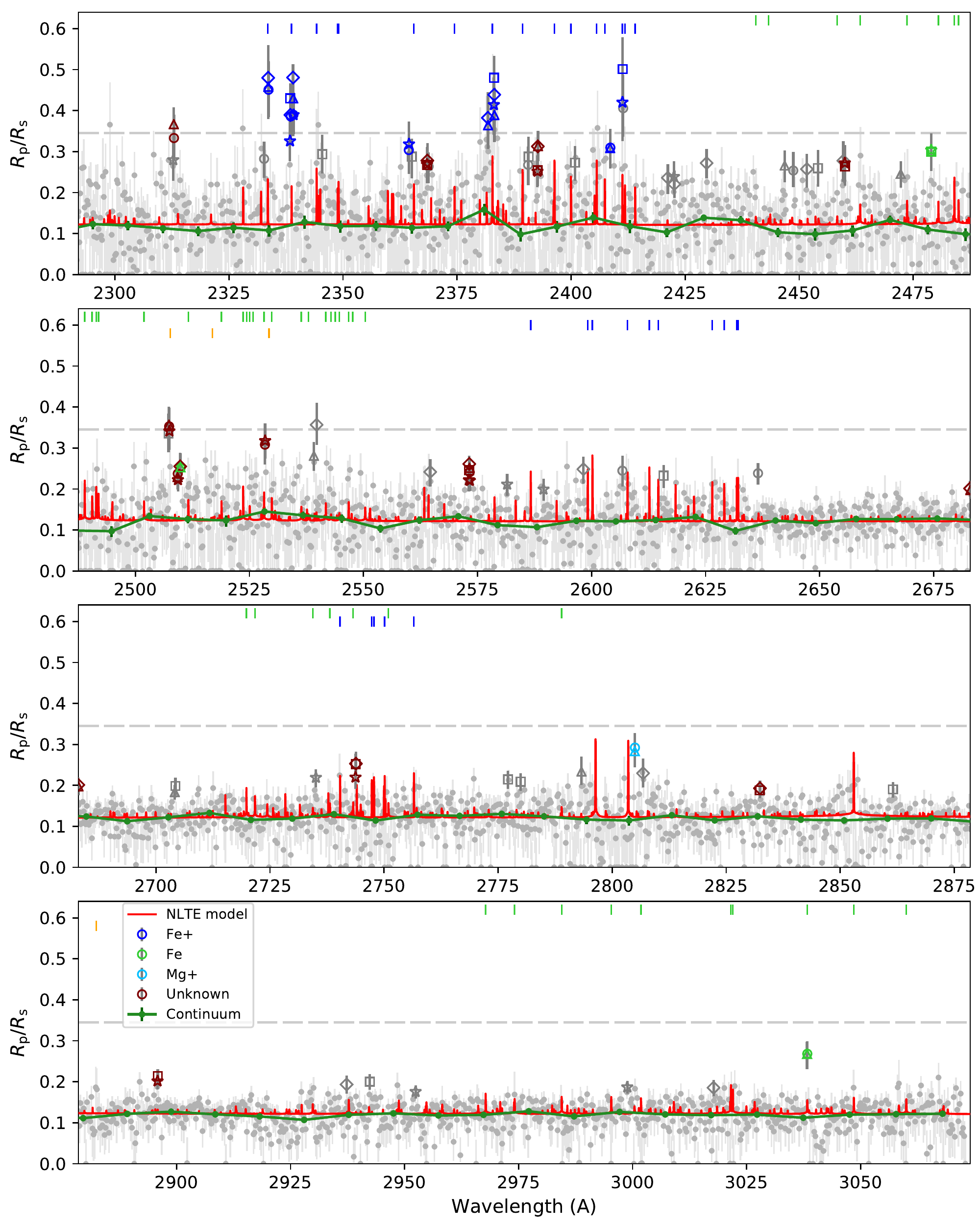}
\caption{
{\twoohnineb} NUV transmission spectrum with outlier absorption
features marked.  The gray dots with error bars denote the observed
spectrum at a resolution of 25\,{\kms} and their 68\% HPD credible
intervals.  The green dots with error bars connected by
a solid line denote the observed spectrum at a resolution of
1000\,{\kms}, which we adopted as the continuum baseline.  The circle,
star, square, diamond, and triangle markers with error bars denote the
$3\sigma$ outliers at each of the shifts (see text).
The colored (see legend) makers denote the outliers that were
identified multiple times at the different shifts, whereas the
dark-gray markers denote the outliers identified at only one shift.
The red curve denotes the theoretical model for this
planet at infinite resolution
(see Section \ref{sec:atmo_model}, for details).  The vertical-line
marks at the top denote all line transitions for a given ion, with
line strengths larger than the detected features in the data (same
color code as in the legend).  We also marked in orange the four
strongest Si line transitions at 2510--2530 and 2880 {\AA}, since they
lay close to two absorption features, though our algorithm cannot
uniquely associate them to Si.} The horizontal dashed line denotes the
size of the Roche lobe in the direction perpendicular to the line of
sight (i.e., as probed by transmission spectroscopy).
\label{fig:spechires}
\end{figure*}

We find that many of the outliers in the {\twoohnineb} NUV data can be
associated with {\feii} absorption lines, at the $P\sb{i} > 68.3$\%
level, three outliers can be associated with {\fei}, and one with
{\mgii}.  Figure \ref{fig:spechires} shows the derived transmission
spectrum at a resolution of 25\,{\kms} highlighting the position of
the deviating data-points, further associating them with the
identified line transitions.  On the top of each panel, we mark the
positions of other lines of the {\fei}, {\feii}, and {\mgii} ions that
should be stronger than the identified ones.  Note that this analysis
implicitly assumes local thermodynamic equilibrium (LTE), while
non-LTE effects are expected to play a significant role in the
atmospheric layers probed by the observations.  Therefore, the marks
on the top of each panel in Figure \ref{fig:spechires} should be
considered to be only indicative.

Figure \ref{fig:spechires} further compares the observed transmission
spectrum to a theoretical model spectrum of {\twoohnineb} (see
Section \ref{sec:atmo_model}).  As expected from the
line-identification analysis, many {\feii} lines present in the model
spectrum correlate with the location of outliers identified as {\feii}
features by the probabilistic approach presented above.

Since our high-resolution binning of 25 {\kms} is larger than the
instrumental resolving element (10 {\kms}, the span of two pixels),
there is an arbitrariness in the selection of the bin edges.
Therefore, to confirm the robustness of our detections, we repeated
the line-identification analysis four more times, shifting the bin
locations by half the instrumental resolving element each time (i.e.,
a total of five iterations shifted by 5~{\kms} from each other); this
is similar to the analysis of \citet{SpakeEtal2018natWASP107bHelium}.
In Figure \ref{fig:spechires} we use a different symbol to mark the
$3\sigma$ outliers at each shift.  The colored markers denote the
lines identified at multiple shifts, whereas the dark-gray markers
denote the lines identified at a single shift location.

The most robust features that we detect are a set of absorption
features between 2330 and 2420 {\AA}, consistent with a {\feii}
band.  Most of these features appear at three or more different
shifts, reassuring us that these are not artifacts resulting from the
data analysis.  We also find a large number of `unknown'
absorption features.  Note that by `unknown', we do not mean that
there are no known line transitions at the location of the outlier,
rather that there is no unique line transition unambiguously
associated with the observed feature.
We do not detect any $3\sigma$ outliers near the {\mgi}
line at 2850~{\AA}.  The {\mgii} doublet at 2800~{\AA} is interesting,
because we only find a recurring absorption feature redward of the
{\mgii} h line, but not near the {\mgii} k line.  This is conflicting,
since the two magnesium lines have similar strengths ({\mgii} h being
slightly weaker), and thus the data should show both features.  Given
that we observe a {\mgii} h feature only at two shifts, and that we do
not detect any feature around the {\mgii} k line, we conclude that
this does not constitute a robust detection of magnesium.


\begin{figure}[t]
\centering
\includegraphics[width=1.0\linewidth, clip, trim=14 0 15 0]{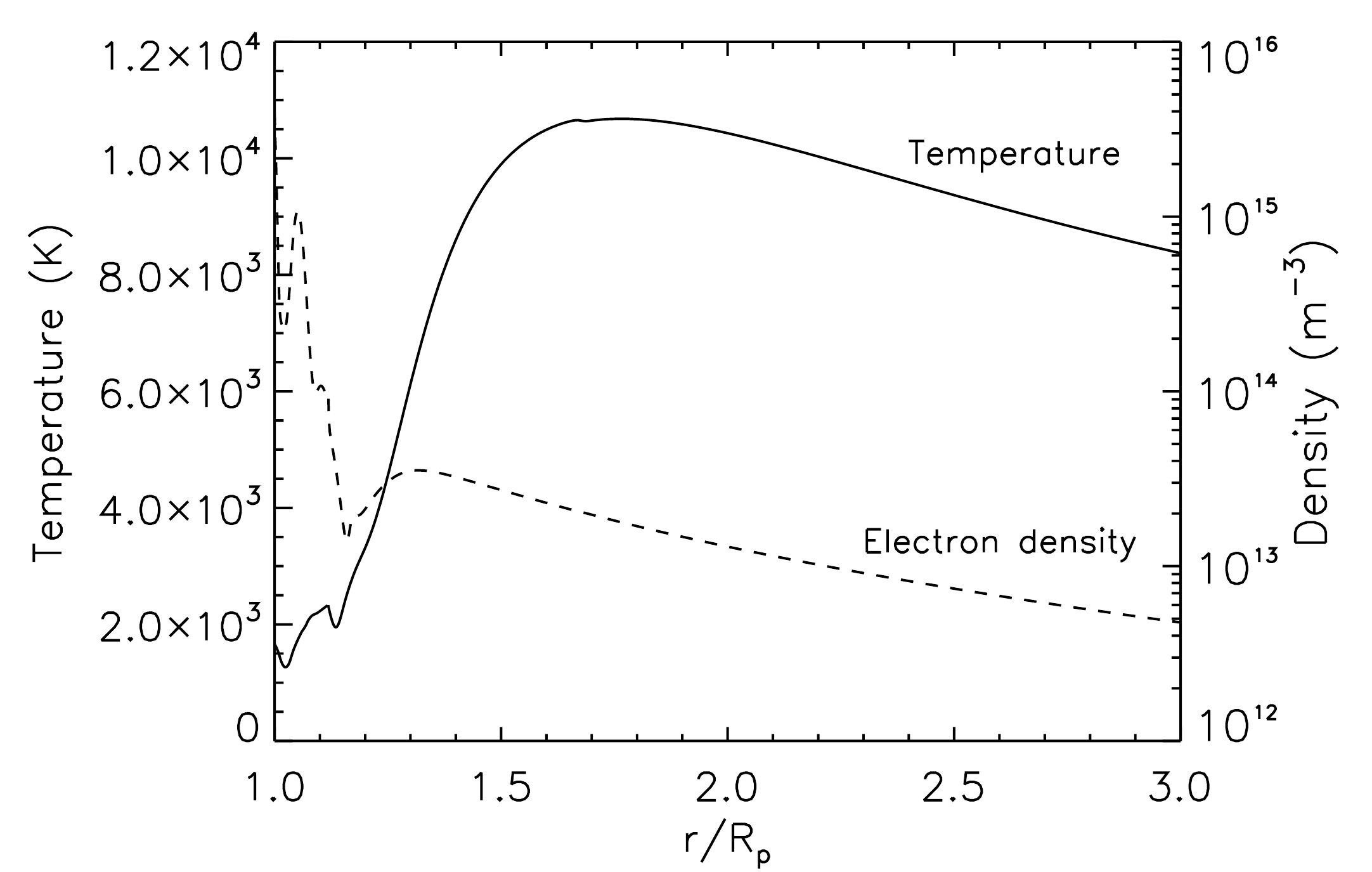}
\includegraphics[width=1.0\linewidth, clip, trim=35 0 20 0]{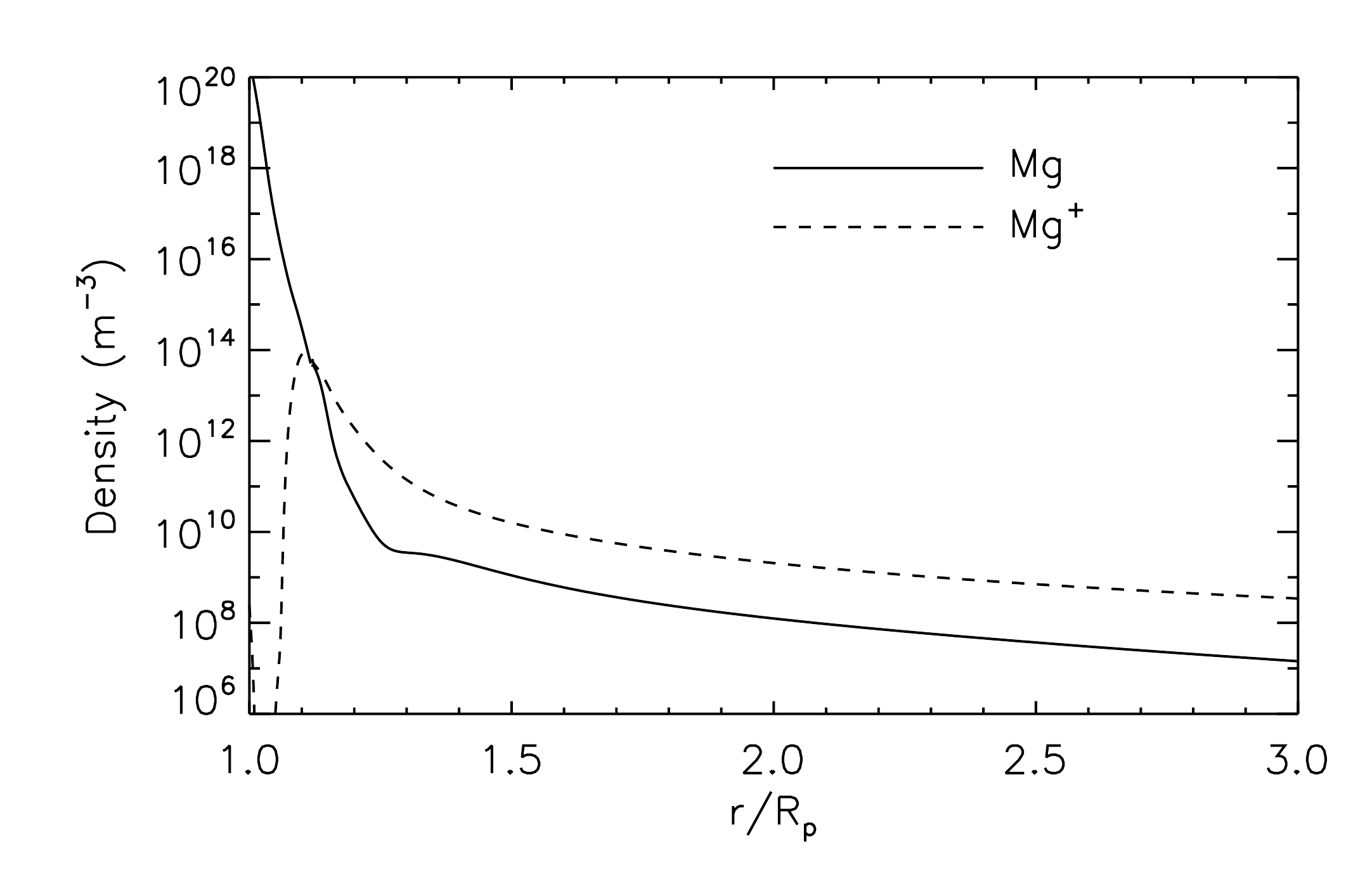}
\caption{
{\twoohnineb} model atmospheric profiles for the temperature and
electron-density (top panel), and the magnesium number density (bottom
panel).  The profiles
here extend from the 1 bar level to the Roche-lobe boundary ($r\approx
3 R\sb{\rm p}$).  Since the sound speed barrier is crossed around the
Roche lobe surface, these profiles are causally separated from the
space above, and thus, unaffected by the physics above the Roche lobe.
Extending the 1D model profiles beyond the Roche lobe would
overestimate the number density, due to lack of 3D effects in our
model; the densities should probably decrease faster with distance
from the planet.  Therefore we capped the densities at the Roche lobe
to produce the transmission spectrum model shown in
Fig.~\ref{fig:spechires}.}
\label{fig:modelatm}
\end{figure}

\subsection{Atmospheric Modeling}
\label{sec:atmo_model}

Here, we simulate theoretical transmission spectra of {\twoohnineb} to
compare against the observed spectra. We adopt the atmospheric
properties (pressure, temperature, and radius) from existing models of
{\twoohnineb} and assume solar abundances for heavy elements.  For the
lower atmosphere, we use the dayside temperature--pressure profile
from \citet{ShowmanEtal2009apj3DCircModel}. For the upper atmosphere
(Figure~\ref{fig:modelatm}), we couple the lower-atmosphere profile to
the model of \citet{KoskinenEtal2013icarHD209458bMetalsEscapeI}. We
then use the spectral-synthesis and radiative-transfer code
{\cloudy} \citep[version 17,][]{FerlandEtal2017rmxaaCloudy} to
calculate ion densities for a static atmosphere and the resulting
transmission spectrum, including the parent neutral species of H,
{\molhyd}, He, Na, Mg and Fe in the model
(Figure~\ref{fig:spechires}).

As the computation of the synthetic transmission spectrum with
{\cloudy} is yet to be published (Young et al., in preparation), we
summarize the procedure here.  Starting from the radial 1D planetary
atmosphere model, we map the altitude, temperature, and abundance
profiles onto concentric, spherically symmetric shells. Next, we
define an array of parallel `transmission' rays, along the
star--observer line of sight, as a function of the planetary impact
parameter.  Then, we track when each ray crosses each atmospheric layer
of the planet to determine the temperature and composition along the
ray path.  We feed these ray profiles into {\cloudy} to produce individual
transmission spectra for each ray.  Finally, we compute the total
integrated transmission spectrum as the sum of the individual spectra,
weighted by the annular area at the given impact parameter, assuming
azimuthal symmetry around the terminator.  We set the density profiles
to zero at the Roche-lobe boundary because 1D atmospheric profiles are
not valid above it.  Therefore, the predicted strength of the
absorption for the lines reaching the Roche-lobe boundary should be
considered to be a lower limit, as the model itself clearly indicates
substantial escape beyond the Roche lobe.  We note that {\cloudy}
includes an extensive database of line transitions and includes
options for both LTE and non-LTE level populations.

The key assumption of this work is that heavy magnesium and iron atoms
with their ions are dragged along by the escaping lighter hydrogen
atoms and protons so that they have roughly solar abundances in the
upper-atmosphere part of the {\cloudy} model.  We used the escape
model of \citet{KoskinenEtal2013icarHD209458bMetalsEscapeI} to
validate this assumption for magnesium (Figure~\ref{fig:modelatm}).
This model does not currently include iron, a much more complex
species, but the mass-loss rate from {\twoohnineb} is predicted to be
sufficiently high to mix iron to high altitudes where {\cloudy} is
used to calculate its ionization balance and transmission in {\fei}
and {\feii} lines.  For magnesium, we show transmission based on the
density profiles predicted by the escape model in
Figure~\ref{fig:magnesium}.  We note that outflow can alter the
ionization balance of metal absorbers and, in particular, enhance
absorption by neutrals, whereas the current version of {\cloudy}
assumes a static atmosphere when calculating the ionization balance.
Therefore, it will eventually also be necessary to study the escape of
iron in detail and our modeling results in Figure \ref{fig:spechires}
should be treated with some caution, although we consider them
perfectly sufficient for a preliminary interpretation of observations.

\section{Results}
\label{sec:results}

The model continuum, based on {\molhyd} and He Rayleigh scattering, is
consistent with the observed baseline (Fig.~\ref{fig:spechires}),
though the scatter does not allow to constrain the Rayleigh slope.

This data alone does not allow us to distinguish between a clear or
cloudy atmosphere. Doing this would require additional NUV
observation, a broader wavelength coverage, or both.  We will address
this issue from a theoretical point of view, in a future work.  From
the optical and infrared observations, it is still under debate
whether {\twoohnineb} has a cloudy or partially cloudy
atmosphere \citep[e.g.,][]{DemingEtal2013apjHD209458b-XO1bWFC3,
MadhusudhanEtal2014apjlH2OabundancesIn3HotJupiters,
SingEtal2016natHotJupiterTransmission,
MacDonaldMadhusudhan2017mnrasRetrievalHD209b,
PinhasEtal2019mnrasHotJupiterSampleRetrieval}.  We acknowledge that
the NUV continuum may also include molecular
absorbers \citep[e.g.,][]{EvansEtal2018ajWASP121b} below the noise
level, but the data do not support the detection of any such absorber.

\begin{figure}[t]
\centering
\includegraphics[width=1.0\linewidth, clip, trim=2 2 2 2]{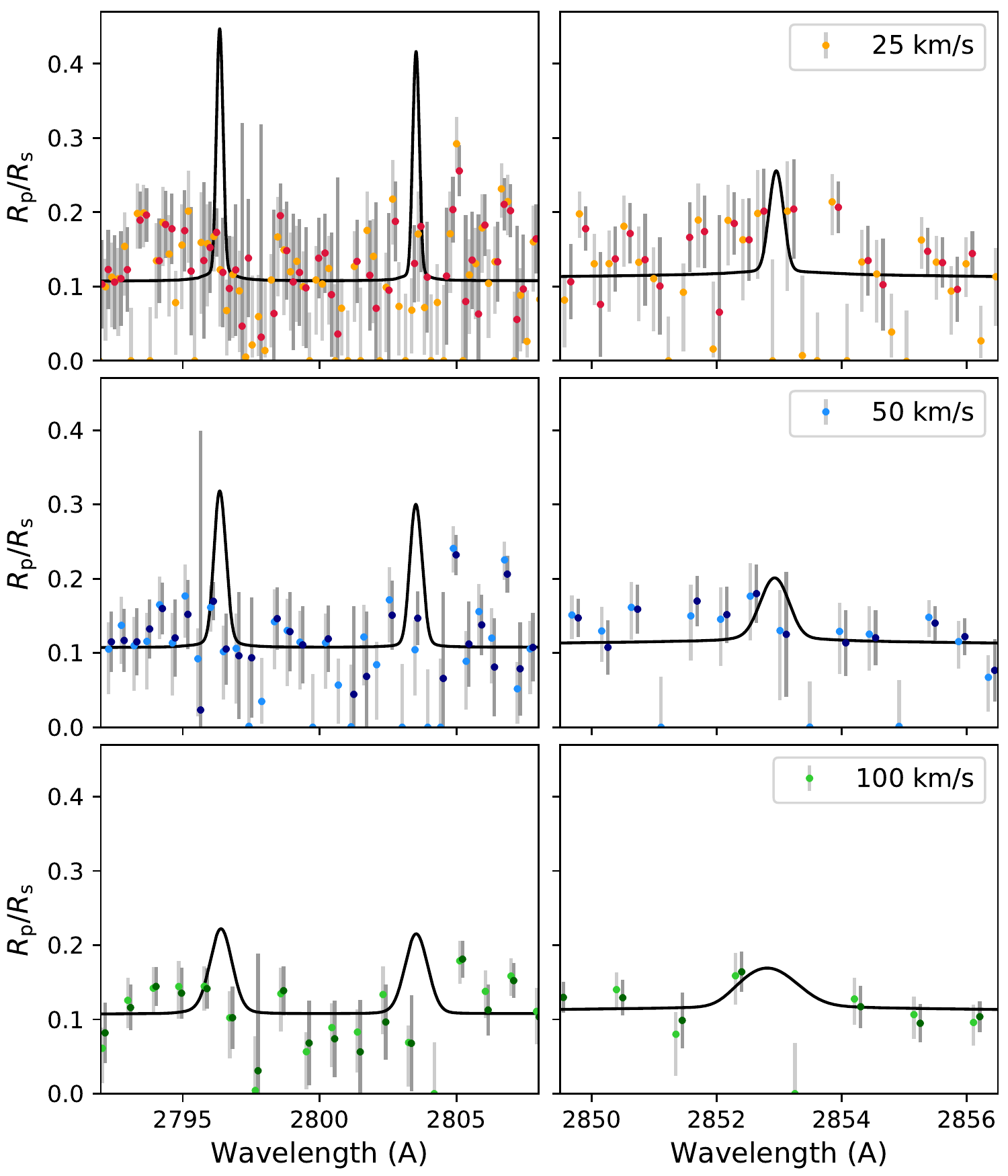}
\caption{
{\twoohnineb} transmission spectrum around the {\mgii}\,h\&k (left)
and {\mgi} (right) resonance lines at three of the four considered
resolving powers (see legend).  The light and dark colored dots denote
the transit depths obtained from the light-curve fit and $AD$ methods,
respectively.  The $AD$ results are shifted horizontally for
visibility. The light and dark gray vertical lines denote the 68\%
HPD credible region of the light-curve fit and the uncertainties of
the $AD$ methods, respectively.  The black solid lines show the
magnesium absorption profiles based on the densities in
Figure~\ref{fig:modelatm} uncapped at the Roche lobe boundary (unlike
in Fig.~\ref{fig:spechires}).  This theoretical spectrum has been
convolved to the resolution given in the legend.  None of the
different analyses detect consistently any significant absorption
feature at the wavelength of the magnesium lines.}
\label{fig:magnesium}
\end{figure}

In line with its low ionization potential, Mg in our models is ionized
near the 1 $\micro$bar pressure level \citep[see Figure 1
in][]{KoskinenEtal2014apjElectrodynamics}, so that {\mgii} and {\feii}
are the dominant magnesium and iron ionization states in the upper
atmosphere.  This is comparable, for example, with the detection of
ionized magnesium and iron in the lower ionosphere of the
Earth \citep{PlaneEtal2015crEarthMesosphereMetals}.

Figure \ref{fig:magnesium} compares the magnesium absorption
signatures predicted by our escape model with our
transmission spectra of {\twoohnineb} for the resonant {\mgii}\,h\&k
lines at about 2800\,\AA\ and {\mgi} line at about 2850\,\AA.  In this
figure, the curve shows the model spectra where the abundances extend
beyond the Roche lobe.  This is an overestimation, because it does not
account for the asymmetric 3D distribution of the material beyond the
planet's Roche lobe.  If the planet had a magnesium overflow beyond
the Roche lobe, both {\mgii} lines should have been observed, given
the assumed magnesium density profile.  However, we do not
consistently detect a significant magnesium signature in the data
(greater than $3\sigma$), at any of the considered resolving powers,
neither from {\mgi} nor {\mgii}.

Our analysis supports the detection of many individual lines
(Figure~\ref{fig:spechires}).  The best match occurs between 2325 and
2420 {\AA}, where the observed spectrum shows robustly detected
absorption features at the location of a dozen of strong {\feii}
lines.  However, the match is not perfect, as there are a few strong
{\feii} lines without a detected counterpart.
More puzzling is that the data does not show robustly detected
features at 2600 {\AA}, where there is another complex of {\feii}
lines that are as strong as those at 2400 {\AA}.
Interestingly, the WASP-121b NUV transmission spectrum reported
by \citet{SingEtal2019ajWASP121bTransmissionNUV}, observed with the
same instrument and detector, also shows a similar trend. There are
numerous other observed absorption features that are too weak to claim
a detection (S/N~$<3\sigma$), or are not consistently detected at
multiple shifts.

Certainly, better calculations of the density profiles, extending
beyond the Roche lobe, are required to constrain the physical
conditions of the upper atmosphere and escape.  Even with these
limitations, however, the model demonstrates that the magnitude of the
detected {\feii} transit depths is consistent with escape models for
the atmosphere of {\twoohnineb}, provided that iron is not removed
from the upper atmosphere by condensation \citep[see,
e.g.,][]{LavvasEtal2014apjElectronDensity}.  This case is similar to
that of WASP-121, where, if the NUV excess absorption is real, it is
most likely caused by {\feii} \citep{SalzEtal2019aaWASP121bSwift}.

\subsection{Comparison with \citet{VidalMadjarEtal2013aaHD209458bSTISnuv}}
\label{sec:vidalmadjar}

Our analysis benefits from advances in data analysis that the
transiting-exoplanet community has experienced over the past decade.
In particular, we applied techniques that were not available or widely
used in the field at the time
of \citet{VidalMadjarEtal2013aaHD209458bSTISnuv}, which allowed us to
analyze this dataset in new light.  In the first place, the
divide-white analysis offers an alternative route to extract the
wavelength-dependent transit depths (as opposed to the $AD(\lambda)$
method).  Second, the Bayesian MCMC analysis allowed us to estimate
more robust transit-depth uncertainties.  Lastly, the wavelength
calibration of the significant radial-velocity shifts over time and
wavelength, might have an impact on the results as well.  Given these
different approaches, and other low-level details in the data
analysis, it can be expected that our results differ from those
of \citet{VidalMadjarEtal2013aaHD209458bSTISnuv}.

Interestingly, our conclusions for the broad-band analysis agree well
with those of \citet{VidalMadjarEtal2013aaHD209458bSTISnuv}, both
finding a baseline transit depth of 1.44\% ($R\sb{\rm p}/R\sb{\rm
s}\sim$\,0.12), where the first two {\HST} visits agree with each other,
and the third visit presents much stronger systematics (in
flux as well as in wavelength calibration).

Regarding the high-resolution analysis of the spectral features, the
detection of the {\mgi} absorption feature
by \citet{VidalMadjarEtal2013aaHD209458bSTISnuv}, absent in our
re-analysis, remains the main discrepancy.  Although we do find
excess-absorption features at transit-depth levels of $\sim$6\%
($R\sb{\rm p}/R\sb{\rm s}\sim$\,0.24) around the location of the {\mgi}
line (see Figure \ref{fig:magnesium}), these are not significant at
the $3 \sigma$ level \citep[consistent with the $2.1\sigma$ result
of][]{VidalMadjarEtal2013aaHD209458bSTISnuv}.  Our further exploration
of the NUV transmission spectrum allowed us to identify a series of
data points with increased absorption deviating more than $3\sigma$
above the continuum, where many of them could be associated to {\feii}
line transitions.

\subsection{Broader Impact: Connection to Lower Atmosphere and FUV Observations}
\label{sec:vidalmadjar}

Although our results resolve the tension given by the previous
detection of {\mgi} and non-detection of {\mgii}, in principle, they
pose a new challenge.  Lower-atmosphere modeling of {\twoohnineb}
indicate that iron and magnesium-bearing cloud condensation curves lie
close enough to each other that if Mg condenses into clouds, then Fe
should also
condense \citep[e.g.,][]{LavvasEtal2014apjElectronDensity}, in direct
contrast with our results.
Recently, using an aerosol microphysics model to directly compute
nucleation and condensation rates, Gao et al. (submitted) find
that magnesium-silicate clouds dominate the aerosol opacity at
planetary equilibrium temperatures of {\twoohnineb}.  The higher
nucleation energy barrier of iron condensates prevents significant
amounts of iron cloud formation, compared to magnesium-silicates, even
though there is as much iron as silicon or magnesium available in the
atmosphere.  Therefore, this microphysics model offers a theoretical
framework consistent with our results.

In relationship to the FUV observations, Shaikhislamov et
al. (submitted) performed three-dimensional hydrodynamic simulations
of the upper atmosphere and circumplanetary environment of
{\twoohnineb}.  They obtained that to fit the 8\% FUV transit-depth
absorption by Si$^{2+}$ observed
by \citet{LinskyEtal2010apjHD209458bMassLoss}, they need to decrease
the silicon abundance to nearly ten times smaller than solar.
Assuming that the clouds are composed mostly of enstatite
(MgSiO\sb{3}) or forsterite (Mg\sb{2}SiO\sb{$4$}), there would be at
least as much magnesium as silicon locked into condensates.  Hence,
with a magnesium abundance of 0.1 times solar, they estimate a
magnesium transit-depth absorption of only 2\%, which is compatible
with the (non-detection) observations presented here.

\section{Conclusions}
\label{sec:conclusions}

We re-analyzed the NUV transmission spectroscopy observations of
{\twoohnineb} presented
by \citet{VidalMadjarEtal2013aaHD209458bSTISnuv} making use of the
improved knowledge and experience in handing systematic noise gathered
in recent years in the exoplanet community. The detection of {\mgi}
and non-detection of {\mgii} presented
by \citet{VidalMadjarEtal2013aaHD209458bSTISnuv} posed a challenge to
the exoplanet atmospheric modeling community, which could not be
solved simply by considering {\mgi} recombination \citep[see also][for
a discussion on this topic]{ShaikhislamovEtal2018apjHD209458b}.
Furthermore, theoretical models predict the presence of a large number
of metallic transition lines in the NUV spectrum.
These considerations led us to this re-analysis.

We employed a two-step analysis: first, identifying the
instrumental systematics from the wavelength-integrated light curves;
and then, extracting the wavelength-dependent transmission spectrum
from the systematics-corrected light curves.  We implemented two
independent spectral analyses, leading to comparable results.
Our transmission spectra indicate the presence of
absorption features extending beyond the Roche-lobe boundary at
wavelengths below $\sim$2500\,{\AA}, with the strongest absorption
concentrating around $\sim$2400\,{\AA}.  A probabilistic
line-identification approach \citep{HaswellEtal2012apjWASP12bNUV}, and
direct comparisons to theoretical models indicate that the detected
absorption features are most probably caused by {\feii} line
transitions.  This requires for iron atoms to be lifted beyond the
planet's Roche lobe; for example, entrained in the hydrodynamic
expansion of the atmosphere by collisions with hydrogen
atoms \citep[e.g.,][]{KoskinenEtal2014apjElectrodynamics}.
We also find no robust evidence of either {\mgi} or {\mgii} absorption
in the planetary upper atmosphere, though, if present at the same
altitudes as {\feii}, the data would enable us to detect it.
There are plausible absorption features around the {\mgii}~h line;
however, the data shows no companion features at a similar shift
around the equally-strong {\mgii}~k line.
In agreement
with \citet{VidalMadjarEtal2013aaHD209458bSTISnuv}, we find that the
NUV continuum absorption is consistent with Rayleigh scattering.

Line broadening from natural and Doppler broadening (FWHM of
$\sim$3~{\kms}), planetary rotational velocity (2~{\kms}), wind
velocities \citep[2~{\kms},][]{SnellenEtal2010natHD209458b}, and
escape velocity \citep[1--10~{\kms},][]{KoskinenEtal2013icarHD209458bMetalsEscapeI} is small
compared to the spectral bin size of 25~{\kms} adopted in our analysis.
However, the radial component of the planet's orbital velocity changes
from the beginning to the end of the transit by
$\sim$30~{\kms} \citep{SnellenEtal2010natHD209458b}, which can make
the absorption features to appear weaker in the analysis.  By
repeating the spectral light-curve fitting at multiple sub-bin shifts
(5~{\kms}), we mitigate ill-posed scenarios where a line might fall
near the edge of a bin.  We will directly account for the orbital
radial-velocity shift in a future study using the high-resolution
cross-correlation technique (in prep.).

By not finding any robust detection of magnesium in the upper
atmosphere, our results resolve the conundrum
presented by \citet{VidalMadjarEtal2013aaHD209458bSTISnuv}.  Furthermore,
recent state-of-the-art microphysics models suggest that cloud
formation can largely favor the formation of magnesium (and therefore,
sequestering) over iron condensates in the lower atmosphere, providing
a plausible theoretical interpretation of our results.

The upper-atmosphere models that we employed are based on an
appropriate hydrodynamic framework and chemical network, but they
adopt a one-dimensional parameterization of the atmosphere, thus
limiting our capability to use the observations to firmly constrain
the physical conditions of the upper atmosphere and escape.  More
complex models, accounting for all relevant physical effects (e.g.,
the three-dimensional dynamics that the material follow beyond the
Roche-lobe boundary) are required to best exploit the already
available and future observations.

Previous \citep{VidalMadjarEtal2013aaHD209458bSTISnuv} and our own
analysis of the observed NUV transmission spectrum of {\twoohnineb}
have revealed challenges to upper-atmosphere models, as well as
results from lower-atmosphere observations.  Having the possibility to
observe both the upper- and lower-atmosphere of an exoplanet
highlights the value of a holistic characterization of exoplanet
atmospheres.  This and other contemporary NUV
studies \citep[e.g.,][]{SingEtal2019ajWASP121bTransmissionNUV} show
that this is possible.

The close-in orbit, large radius, and bright host star of
{\twoohnineb} give this planet a pivotal role for exoplanet
atmospheric characterization.  Few other planets will ever enable such
precise measurements of their upper- and lower-atmosphere properties.
Thus, a better understanding of the upper-atmosphere of this
important target would prove to be particularly useful, especially
before the imminent launch of the {\it James Webb Space Telescope}.
While we robustly detected several absorption features matching
{\feii} transition lines on {\twoohnineb} (at a S/N~$>3\sigma$), there
are many other weaker features in the spectrum.  If the features that
we see are of astrophysical nature, further NUV observations could
potentially uncover from dozens to hundreds of additional {\fei} and
{\feii} features, and place stronger constraints to the strength of
the {\mgi} and {\mgii} lines (if present at all).  Data of the
necessary quality could only be currently obtained with {\HST}.

A dedicated mission for UV spectroscopic observations, such as the
Colorado Ultraviolet Transit
Experiment \citep[CUTE,][]{FlemingEtal2018jatisCUTE}, is therefore an
efficient way to collect NUV observations to prove or disprove the
presence of magnesium in the upper atmosphere.
The combined CUTE and HST data will enable a firm interpretation of
the results, and provide solid constraints to lower and upper atmosphere
modeling.

This work demonstrates the need for comparative analyses of archival
data sets with full transparency and for repeated observations of the
same target aiming at improving data quality, transit light curve
coverage, and robustness of the results against instrumental and
astrophysical noise.

\acknowledgments

We thank the anonymous referee for his/her time and valuable comments.
We thank Peter Gao and Ildar Shaikhislamov for valuable discussions
that helped to place our results in a broader context. We thank
contributors to the Python Programming Language and the free and
open-source community (see Software section below).  This research has
made use of NASA's Astrophysics Data System Bibliographic Services.
We drafted this article using the AASTeX6.2 latex
template \citep{AASteamHendrickson2018aastex62}, with further style
modifications available
at \href{https://github.com/pcubillos/ApJtemplate}
{https://github.com/pcubillos/ApJtemplate}.  A.G.S. and
L.F. acknowledge financial support from the FFG projects 859718 and
865968.
M.E.Y. and L.F. acknowledge funding from the ``Innovationfonds
Forschung, Wissenschaft und Gesellschaft'' project of the Austrian
Academy of Sciences entitled ``Towards Extra-Terrestrial Habitats
(TETH)''.
C.A.H. gratefully acknowledges financial support from STFC under grant
ST/P000584/1.
Based on observations made with the NASA/ESA Hubble Space Telescope,
obtained at the Space Telescope Science Institute, which is operated
by the Association of Universities for Research in Astronomy, Inc.,
under NASA contract NAS 5-26555.  These observations are associated
with program \#11576.

\software{MC3\footnote{\href{https://github.com/pcubillos/mc3}
                       {https://github.com/pcubillos/mc3}}
\citep[][]{CubillosEtal2017apjRednoise},
Limb-darkening\footnote{\href{https://github.com/nespinoza/limb-darkening}
                           {https://github.com/nespinoza/limb-darkening}}
 \citep[][]{EspinozaJordan2015mnrasLimbDarkeningI},
{\cloudy} \citep{FerlandEtal2017rmxaaCloudy},
\textsc{Numpy} \citep{vanderWaltEtal2011numpy},
\textsc{SciPy} \citep{JonesEtal2001scipy},
\textsc{Matplotlib} \citep{Hunter2007ieeeMatplotlib},
\textsc{AstroPy} \citep{Astropy2013aaAstroPy},
\textsc{IPython} \citep{PerezGranger2007cseIPython},
AASTeX6.2 \citep{AASteamHendrickson2018aastex62},
ApJtemplate (\href{https://github.com/pcubillos/ApJtemplate}
                  {https://github.com/pcubillos/ApJtemplate}), and
\textsc{bibmanager}\footnote{\href{https://pcubillos.github.io/bibmanager}
                             {https://pcubillos.github.io/bibmanager}}
 \citep[][]{Cubillos2019zndoBibmanager}.
}

\bibliography{HD209458b_NUV}

\appendix

\section{Supplementary Figures}
\label{sec:appendix}
\renewcommand{\thefigure}{A\arabic{figure}}
\setcounter{figure}{0}

\begin{figure*}[b]
\centering
\includegraphics[width=\linewidth, clip, trim=2 47 2 3]{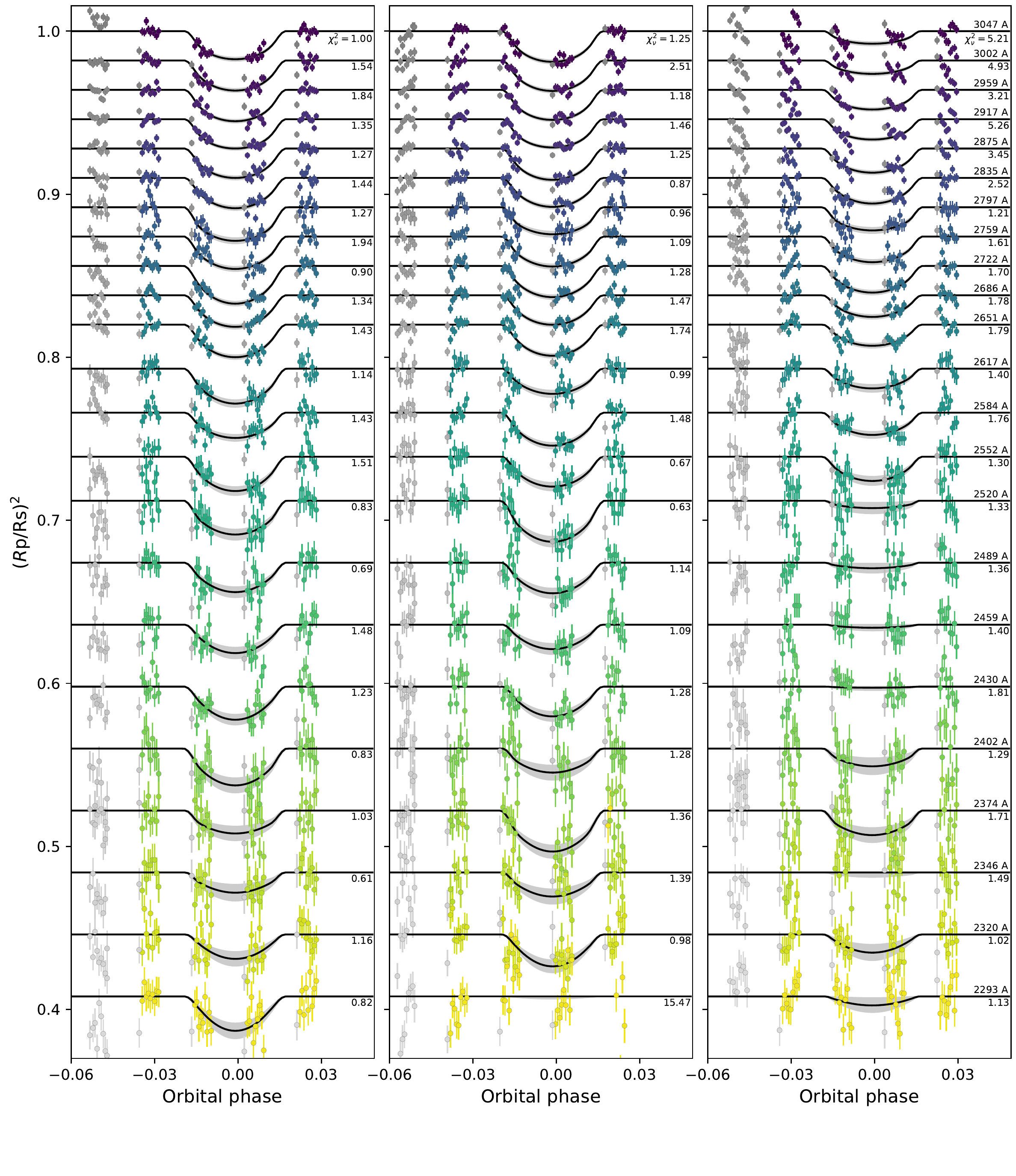}
\caption{
Light-curve observations (markers with error bars) and best-fit model
(black curves) for each visit (left to right panels) and spectral
order (top to bottom curves). The shaded gray area denote the span of
models for the 0.68-quantile from the MCMC posterior distribution. The numbers
below each model show the reduced chi-squared ($\chi_\nu^2$) for each
light curve.  The numbers above each model on the right panel show the
central wavelength of each spectral order.}
\label{fig:appendix_lc}
\end{figure*}

\begin{figure*}[b]
\centering
\includegraphics[width=0.49\linewidth, clip, trim=7 0 10 0]{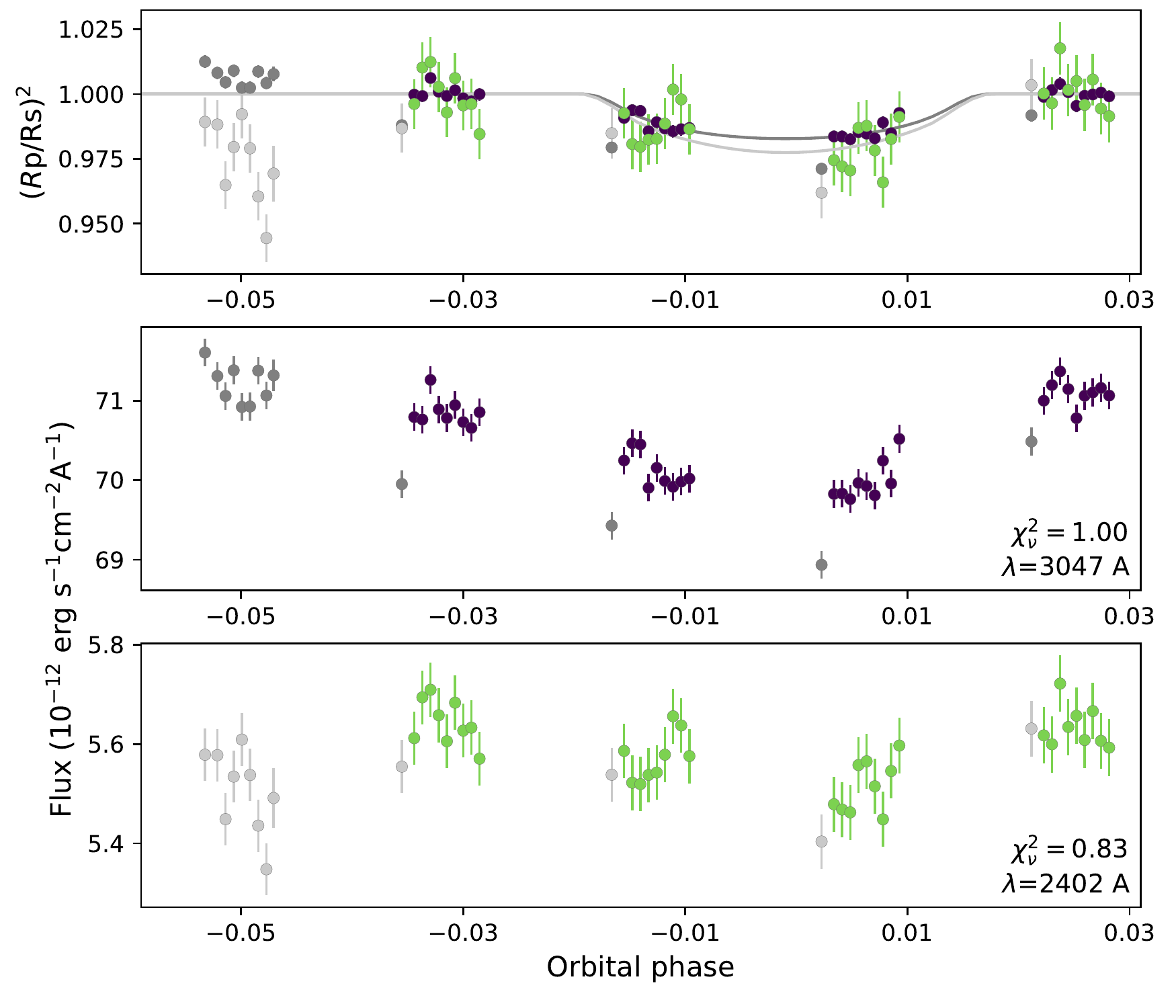}
\includegraphics[width=0.49\linewidth, clip, trim=7 0 10 0]{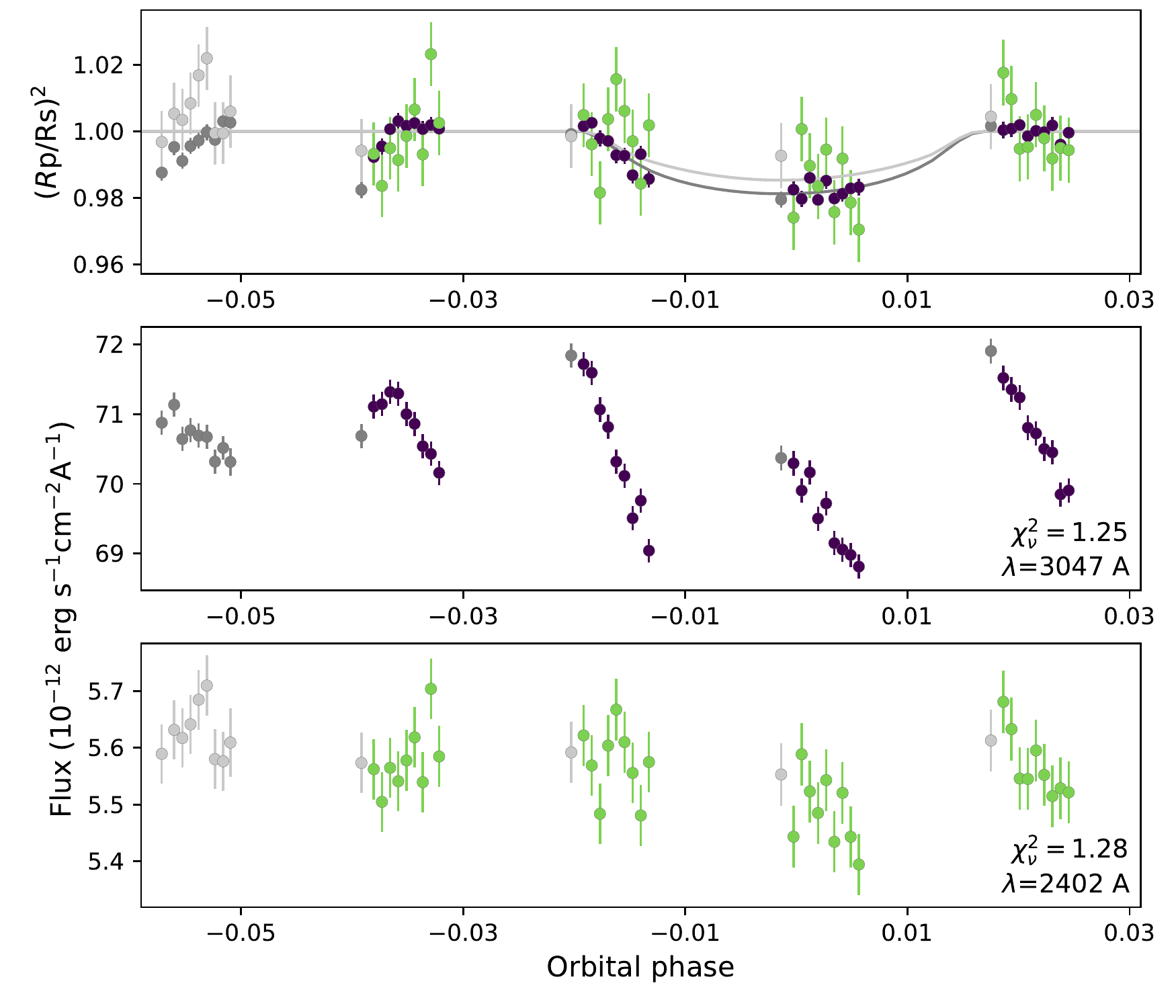}
\includegraphics[width=0.49\linewidth, clip, trim=7 0 10 0]{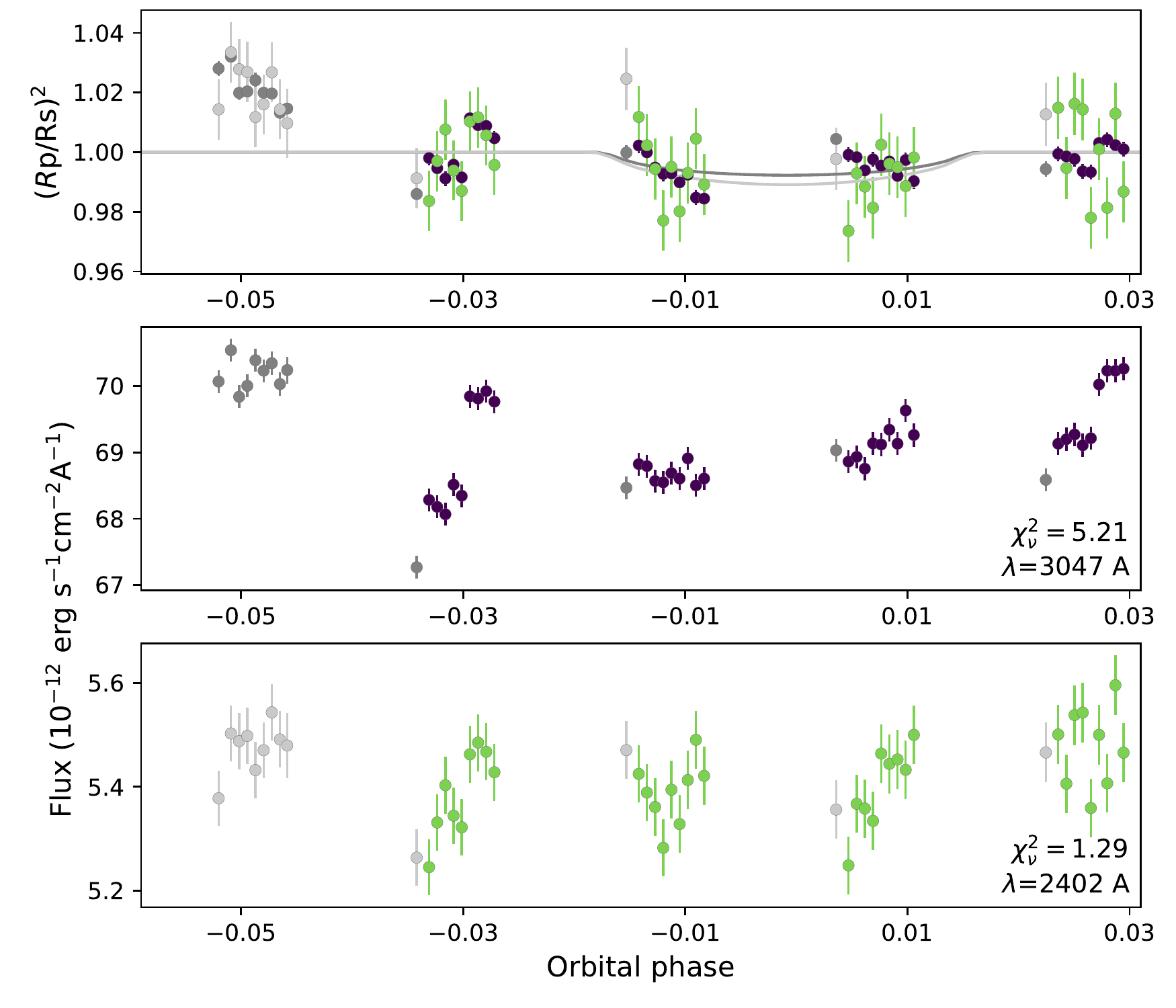}
\caption{
Sample order-integrated light curves for visit 1 (top-left set of
panels), visit 2 (top-right set of panels), and visit 3 (bottom set of
panels).  For each visit, the top panel shows the
systematics-corrected light curve for the 1st ($\lambda=3047$~A, dark
colors) and 19th ($\lambda=2402$~A, light colors) order, whereas the
middle and bottom panels show the raw integrated-flux light curves for
the 1st and 19th orders, respectively.}
\label{fig:appendix_systmatics}
\end{figure*}

\end{document}